\documentclass[aps,english,prl,amsmath,amsfonts,amssymb,superscriptaddress,onecolumn,showpacs,citeautoscript,preprint]{revtex4-1}
\usepackage[T1]{fontenc}
\usepackage[latin9]{inputenc}
\usepackage{amsmath}
\usepackage{amssymb}
\usepackage{wasysym}
\usepackage{esint}
\usepackage{graphicx}
\usepackage{times}
\usepackage{nicefrac}
\usepackage{appendix}
\usepackage{textcase}
\usepackage{subfigure}
\usepackage{empheq}
\usepackage{color}
\usepackage{leftidx}
\usepackage{makecell}
\usepackage{stackrel}
\makeatletter
\@ifundefined{textcolor}{}
{%
 \definecolor{BLACK}{gray}{0}
 \definecolor{WHITE}{gray}{1}
 \definecolor{RED}{rgb}{1,0,0}
 \definecolor{GREEN}{rgb}{0,1,0}
 \definecolor{BLUE}{rgb}{0,0,1}
 \definecolor{CYAN}{cmyk}{1,0,0,0}
 \definecolor{MAGENTA}{cmyk}{0,1,0,0}
 \definecolor{YELLOW}{cmyk}{0,0,1,0}
}

\renewcommand{\vec}[1]{\mathbf{#1}}
\renewcommand{\Re}{\operatorname{Re}}
\renewcommand{\Im}{\operatorname{Im}}

\newcommand{\sym}{\operatorname{sym}}

\renewcommand{\b}{\beta}

\newcommand{\add}[1]{\if\a\b{{\color{red} #1}}\else{#1}\fi}

\renewcommand{\eqref}[1]{Eq.~\ref{eq:#1}}

\newcommand{\Eqref}[1]{Equation~\ref{eq:#1}}
\newcommand{\figref}[1]{Fig.~\ref{fig:#1}}
\newcommand{\Figref}[1]{Figure~\ref{fig:#1}}

\def\ro{\vec{x}}
\def\rs{\vec{y}}
\newcommand{\mat}[1]{{#1}}
\newcommand{\matvec}[1]{{#1}}

\newcommand{\trace}[1]{{\rm Tr} \left[ #1 \right]}

\newcommand{\Tr}{\text{Tr }}

\makeatother

\usepackage{babel}
\begin{document}
\title{Temperature control of thermal radiation from heterogeneous bodies}

\author{Weiliang Jin}
\affiliation{Department of Electrical Engineering, Princeton University, Princeton, NJ 08544, USA}
\author{Athanasios G. Polimeridis}
\affiliation{Skolkovo Institute of Science and Technology, Moscow, Russia}
\author{Alejandro W. Rodriguez}
\affiliation{Department of Electrical Engineering, Princeton University, Princeton, NJ 08544, USA}

\begin{abstract}
  We demonstrate that recent advances in nanoscale thermal transport
  and temperature manipulation can be brought to bear on the problem
  of tailoring thermal radiation from compact emitters. We show that
  wavelength-scale composite bodies involving complicated arrangements
  of phase-change chalcogenide (GST) glasses and metals or
  semiconductors can exhibit large emissivities and partial
  directivities at mid-infrared wavelengths, a consequence of
  temperature localization within the GST. We consider multiple object
  topologies, including spherical, cylindrical, and mushroom-like
  composites, and show that partial directivity follows from a
  complicated interplay between particle shape, material dispersion,
  and temperature localization. Our calculations exploit a recently
  developed fluctuating--volume current formulation of electromagnetic
  fluctuations that rigorously captures radiation phenomena in
  structures with both temperature and dielectric inhomogeneities.
\end{abstract}

\maketitle

The ability to control thermal radiation over selective frequencies
and angles through complex materials and nanostructured
surfaces~\cite{greffet2007coherent} has enabled unprecedented advances
in important technological areas, including remote temperature
sensing~\cite{masuda1988emissivity}, incoherent
sources~\cite{ilic2014thermal,rinnerbauer2014metallic}, and
energy-harvesting~\cite{fan2014photovoltaics,bermel2011tailoring,florescu2007improving}.
Recent progress in the areas of temperature management and thermal
transport at sub-micron scales can play a significant (and largely
unexplored) role in the design of specially engineered radiative
structures that combine both photonic and phononic design
principles~\cite{cahill2003nanoscale,cahill2014nanoscale}.

\begin{figure}[t!]
\begin{center}
\includegraphics[width=0.55\columnwidth]{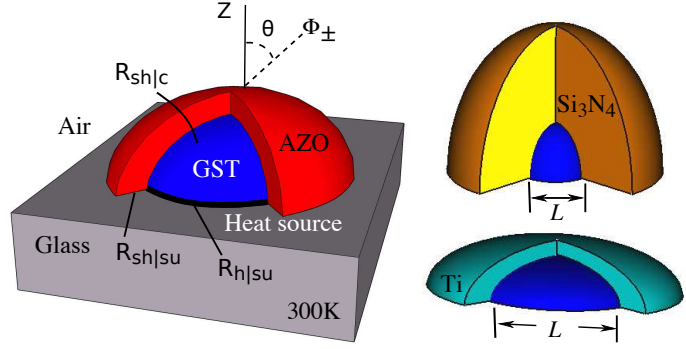}
\caption{Schematic of representative composite bodies comprising GST
  (blue) hemispheroids coated with Ti (green), AZO (red), or
  Si$_3$N$_4$ (orange) shells, and resting on a low-index, transparent
  substrate in contact with a heat reservoir at 300~K. The GST is
  heated from below by a conductive 2D material (e.g. a
  carbon-nanotube wall or graphene sheet), leading to temperature
  gradients within the structure. The presence of boundary resistance
  at material interfaces is captured by effective (intrinsic and
  contact) thermal resistances $R_\mathrm{th}$.}
\label{fig:fig1}
\end{center}
\end{figure}

In this letter, we describe a fluctuating--volume current (FVC)
formulation of electromagnetic (EM) fluctuations that enables fast and
accurate calculations of thermal radiation from complex structures
with non-uniform temperature and dielectric properties. We demonstrate
that when selectively heated, wavelength-scale composite
bodies---complicated arrangements of phase-change materials and metals
or semiconductors---can exhibit large temperature gradients and
partially directed emission at infrared wavelengths. For instance,
micron-scale chalcogenide (GST) hemispheroids coated with titanium or
silicon-nitride shells [\figref{fig1}] and resting on a low-index,
transparent substrate can exhibit large emissivity and $\gtrsim 80\%$
partial directivity---redirecting light away from the metallic or
toward the semiconducting shell---when heated to $870$~K by highly
conductive 2D materials at the GST--substrate interfaces. The
interplay of geometry and temperature localization allows such
composite infrared thermal antennas to not only enhance but also
selectively emit and absorb light in specific directions. We show that
other designer bodies, including mushroom-like particles and coated
cylinders [\figref{fig3}], can also exhibit large partial directivity,
in contrast to situations involving homogeneous bodies or uniform
temperature distributions which lead to nearly isotropic emission. Our
predictions are based on accurate numerical solutions of the
conductive heat equation and Maxwell's equations, which not only
incorporate material dispersion but also account for the existence of
thermal and dielectric gradients at the scale of the EM wavelength,
where ray optical descriptions are inapplicable.

Attempts to obtain unusual thermal radiation patterns have primarily
relied on Bragg scattering and related interference effects in
nanostructured surfaces~\cite{greffet2007coherent}, including photonic
gratings~\cite{de2012conversion,wang2014thermal,florescu2007improving},
metasurfaces~\cite{greffet2002coherent,marquier2004coherent,Joulain05,hesketh1988polarized,narayanaswamy2005thermal,marquier2015metallic,kleiner2012acrobatics,ribaudo2013highly},
multilayer
structures~\cite{kollyukh2003thermal,ben2004thermal,drevillon2011far,wang2011direct},
and sub-wavelength
metamaterials~\cite{lee2008coherent,fu2009thermal,Liu11,bermel2011tailoring,liu2011taming}. Related
ideas can also be found in the context of fluorescence emission, where
directivity is often achieved by employing metallic objects
(e.g. plasmonic antennas) to re-direct emission from individual
dipolar emitters via
gratings~\cite{curto2010unidirectional,kosako2010directional} or by
localizing fluorescent molecule(s) to within some region in the
vicinity of an external
scatterer~\cite{teperik2011numerical,thomas2004single,li2007fdtd,vandenbem2009fluorescence,vandenbem2010controlling,mohammadi2008gold}. Matters
become complicated when the emission is coming from random sources
distributed within a wavelength-scale object, as is the case for
thermal radiation, because the relative contribution of current
sources to radiation in a particular direction is determined by both
the shape and temperature distribution of the object. Optical antennas
have recently been proposed as platforms for control and design of
narrowband emitters~\cite{schuller2009optical,novotny2011antennas},
though predictions of large directivity continue to be restricted to
periodic structures.  While there is increased focus on the study of
light scattering from subwavelength particles and microwave antennas
(useful for radar detection~\cite{balanis2005antenna},
sensing~\cite{taminiau2007lambda}, and color
routing~\cite{alavi2012color,shegai2011bimetallic}), similar ideas
have yet to be translated to the problem of thermal radiation from
compact, wavelength-scale objects, whose radiation is typically
quasi-isotropic~\cite{greffet2007coherent}. Here, we show that
temperature manipulation in composite particles could play an
important role in the design of coherent thermal emitters.

Temperature gradients can arise near the interface of materials with
highly disparate thermal
conductivities~\cite{cahill2014nanoscale}. Although often negligible
at macroscopic scales~\cite{balandin2011thermal}, recent experiments
reveal that the presence of thermal boundary resistance
~\cite{reifenberg2008impact,marconnet2013thermal} (including intrinsic
and contact resistance~\cite{stevens2007effects}) in nanostructures
together with large dissipation can enable temperature localization
over small distances~\cite{balandin2011thermal}. Such temperature
control has been recently investigated in the context of metallic
nanospheres immersed in fluids~\cite{merabia2009critical}, graphene
transistors~\cite{islam2013role}, nanowire resistive
heaters~\cite{yeo2014single}, AFM tips~\cite{king2013heated}, and
magnetic contacts~\cite{petit2012understanding}. With the exception of
a few high--symmetry structures, e.g.
spheres~\cite{dombrovsky2000thermal} and planar
films~\cite{wang2011direct}, however, calculations of thermal
radiation from wavelength-scale bodies have been restricted to
uniform-temperature operating conditions, exploiting Kirchoff's
law~\cite{Luo04:thermal,wang2011direct} to obtain radiative emission
via simple scattering calculations. The presence of temperature and
dielectric inhomogeneities in objects with features at the scale of
the thermal and EM wavelengths call for alternative theoretical
descriptions.

\emph{Formulation.---} In what follows, we present a brief derivation
of our FVC formulation of thermal radiation, with validations and
details of its numerical implementation described in a separate
manuscript~\cite{polimeridis2015fluctuating}. Our starting point is
the VIE formulation of EM scattering~\cite{polimeridis2014stable},
describing scattering of an incident, 6-component electric ($\vec{E}$)
and magnetic ($\vec{M}$) field $\phi_{\rm inc} = (\vec{E}; \vec{H})$
from a body described by a spatially varying $6\times 6$
susceptibility tensor $\chi(\ro)$. (For convenience, we omit the
frequency $\omega$ dependence of material properties, currents,
fields, and operators, and also define $\chi$ to be the susceptibility
relative to the background medium.) Given a 6-component electric
($\vec{J}$) and magnetic ($\vec{M}$) dipole source $\sigma =
(\vec{J};\vec{M})$, the incident field is obtained via a convolution
($\star$) with the $6\times 6$ homogeneous Green's function (GF) of
the ambient medium $\Gamma(\ro,\rs)$, such that
$\phi_{\mathrm{inc}}=\Gamma\star\sigma=\int d^3 \rs
\Gamma(\ro,\rs)\sigma(\rs)$. Exploiting the volume equivalence
principle~\cite{polimeridis2014stable}, the unknown scattered fields
$\phi_{\rm sca} = \Gamma \star \xi$, can also be expressed via
convolutions with $\Gamma$, except that here $\xi=-i\omega\chi\phi$
are the (unknown) bound currents in the body, related to the total
field inside the body $\phi=\phi_\mathrm{inc} + \phi_\mathrm{sca}$
through $\chi$. Writing Maxwell's equations in terms of the incident
and induced currents,
\begin{equation}
  \xi + i \omega \chi (\Gamma \star \xi)= -i \omega \chi (\Gamma
  \star \sigma),
  \label{eq:VIEop}
\end{equation}
one obtains $\xi$ in terms of the incident source $\sigma$.

Consider a Galerkin discretization of \eqref{VIEop} via expansions of
the current sources $\sigma(\ro)=\sum_n s_n b_n(\vec{x})$ and
$\xi(\ro)=\sum_n x_n b_n(\vec{x})$ in a convenient, orthonormal basis
$\{b_n\}$ of $N$ 6-component vectors, with vector coefficients $s$ and
$x$, respectively. The resulting matrix expression has the form
$x+s=Ws$, where $(\mat{W}^{-1})_{m,n}= \langle b_m,b_n + i \omega \chi
(\Gamma \star b_n) \rangle$ is known as the VIE matrix and $\langle ,
\rangle$ denotes the standard conjugated inner product.  Poynting's
theorem implies that the far-field radiation flux $\Phi = \frac{1}{2}
\Re \int d^{3}\vec{x}\, (\vec{E}^*\times \vec{H})$ due to $\sigma$,
\begin{align*}
  \Phi_\sigma &=-\frac{1}{2}\Re \xi^{*}\phi =-\frac{1}{2}\Re
  (\xi+\sigma)^{*}\Gamma\star(\xi+\sigma)\\&=-\frac{1}{2}
  (x+s)^{*}\sym G(x+s) =-\frac{1}{2} s^{*}W^{*}\sym GWs \\ & =
  -\frac{1}{2} \Tr \left[(s^* s) W^* \sym G W\right]
\end{align*}
where $s^*s$ and $G$ are both $N\times N$ matrices and
$\mat{G}_{mn}=\langle b_m,\Gamma\star b_n~\rangle$ are the elements of
the so-called ``Green'' matrix. Thermal radiation from such a body
follows from the cumulative flux contributions of a collection of
incoherent sources distributed throughout its
volume~\cite{polimeridis2015fluctuating}, obtained by a thermodynamic,
ensemble-average $\Phi = \langle \Phi_\sigma \rangle$ over all
$\sigma$ and polarizations. It follows that the total radiation is
given by:
\begin{equation}
  \Phi = -\frac{1}{2} \trace{ \mat{C} \mat{W}^\ast \sym{ \mat{G}} \mat{W}},
\label{eq:Utotal}
\end{equation}
where we have defined the current--current correlation matrix $C$,
whose elements $\mat{C}_{mn}=\langle \matvec{s_m}
\matvec{s_n}^\ast\rangle = \int \int d^3\ro d^3\rs \, b_{m}^\ast(\ro)
\langle \sigma(\ro) \sigma^\ast(\rs) \rangle b_{n}(\rs)$. The
correlation functions satisfy a well-known fluctuation--dissipation
theorem (FDT)~\cite{landau:stat2}, $\langle \sigma_i(\ro,\omega)
\sigma_j^\ast(\rs,\omega) \rangle = \frac{4}{\pi} \omega\Im
\chi(\ro,\omega) \Theta(\ro,\omega)\delta(\ro-\rs)\delta_{ij}$,
relating current fluctuations to the dissipative $\sim \Im \chi$ and
thermodynamic properties of the underlying materials. Here,
$\Theta(\vec{x},\omega) = \hbar \omega / (e^{\hbar \omega/k_{\rm B}
  T(\vec{x})} - 1)$ is the Planck distribution at the local
temperature $T(\vec{x})$~\cite{rodriguez2013fluctuating}. In addition
to the total flux, it is also desirable to obtain the angular
radiation pattern from bodies, which can be straightforwardly obtained
by introducing the far-field Green matrix $\mat{G}_{\infty}^{\rm
  E\ast}=\langle b_m, Q\Gamma_\infty^{\rm E} \star b_n \rangle$, based
on the $3\times 6$ GF $\Gamma_\infty^{\rm E}(\ro,\rs)$ which maps
currents to far-field electric fields and the $3\times 3$ tensor $Q$
mapping Cartesian to spherical coordinates (azimuthal and polar
angles)~\cite{Balanis97}. Following a similar derivation, the angular
radiation flux in a given direction will be given by:
\begin{equation}
  \label{eq:Uangle}
  U = \frac{k^2 Z}{2(4\pi)^2}\trace{\mat{C}\,\mat{W}^\ast
    (\mat{G}_{\infty}^{\rm E\ast}\mat{G}_{\infty}^{\rm E}) \mat{W}},
\end{equation}
where $k=\omega/c$ and $Z=\sqrt{\mu_0/\varepsilon_0}$ is the wave
impedance of the background medium. \Eqref{Uangle} can be employed to
calculate emission from arbitrarily shaped bodies with spatially
varying dielectric and temperature properties---unlike previous
scattering-matrix and surface-integral equation formulations of
thermal radiation~\cite{rodriguez2013fluctuating}, the FVC scattering
unknowns are volumetric currents. Furthermore, when considered as a
numerical method as we do below, the corresponding basis functions can
be chosen to be localized ``mesh'' elements~\cite{Johnson11:review},
allowing resolution to be employed where
needed~\cite{polimeridis2015fluctuating}.

%Furthermore, although not entirely obvious, both trace formulas are
%susceptible to iterative, fast-trace computations since they involve
%the Frobenium norm of low-rank matrices~[ref].

%In what follows, \eqref{Uangle} is employed to calculate the
%thermal flux from a variety of complex structures with spatially
%varying temperature and dielectric profiles.

%TODO: For more detailed description of the computation
%convergence and evaluation of [eq], the reader is referred to [ref].

\begin{figure*}[t!]
\begin{center}
\includegraphics[width=1\columnwidth]{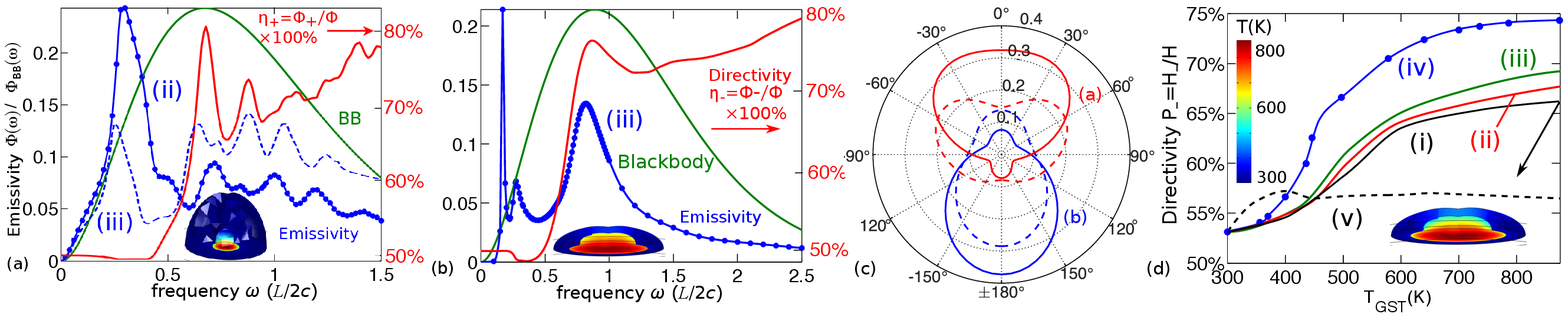}
\caption{Spectral emissivity $\epsilon(\omega) =
  \Phi(\omega)/\Phi_\mathrm{BB}(\omega)$ (blue dots) from composite
  bodies comprising Ge$_2$Sb$_2$Te$_5$ (GST) hemispheroids coated with
  either (a) Si$_3$N$_4$ or (b) Ti shells, under heating scenarios
  (ii) and (iii), respectively (see text). The dashed blue line in (a)
  shows $\epsilon$ for the Si$_3$N$_4$ structure under heating
  scenario (iii). Both structures rest on a low-index substrate (shown
  schematically in \figref{fig1}) and are heated from the
  GST--substrate interface by a 2D thin-film conductor, reaching an
  interface temperature $T_\mathrm{GST} = 870$~K; the temperature
  profiles $T(\vec{x})$ are shown as insets. $\epsilon$ is defined as
  the ratio of the thermal flux $\Phi(\omega)$ from each body
  normalized to the flux
  $\Phi_{\mathbf{BB}}=\frac{A}{4\pi^2}(\omega/c)^2\Theta(\omega,T)$
  from a corresponding black body of same area $A$ and uniform
  temperature $T=870$~K (green lines, arbitrary units). Also shown are
  the partial directivities $\eta_{\pm} = \Phi_{\pm} / \Phi$ (red
  line), defined as the ratios of the outgoing flux into the
  upper/lower hemisphere $\Phi_{\pm} = 2\pi \int_{\mp \pi/2}^{\pi\mp
    \pi/2} d\theta\, \Phi(\omega,\theta)$ to the total flux, where
  $\theta$ is defined in \figref{fig1}. (c) Angular radiation
  intensity $U(\theta)$ for the structures in (a) (solid red line) or
  (b) (solid blue line) as well as under heating scenario (v),
  corresponding to uniform temperature distributions (dashed
  lines). (d) Total (frequency integrated) downward partial
  directivity $P_-=H_{-}/H$ as a function of $T_\mathrm{GST}$, where
  $H_- = \int_0^\infty d\omega\, \Phi_-(\omega)$ and $H=\int_0^\infty
  d\omega\, \Phi(\omega)$, for the Ti structure under different
  heating conditions, corresponding to multiple degrees of temperature
  localization in the GST (see text).}
\label{fig:fig2}
\end{center}
\end{figure*}

% with GST long (short) semi-axes $830$nm ($330$nm) and overall long
%(short) semi-axes $1.16\mu$m ($460$nm), corresponding to a metal
%thickness $\approx 400$nm.

\emph{Results.---} In what follows, we explore radiation from composite bodies
comprised of chalcogenide Ge$_2$Sb$_2$Te$_5$ (GST) alloys and metals
or semiconductors. To begin with, we consider micron-scale GST
hemispheroids coated with either titanium (Ti) or silicon-nitride
(Si$_3$N$_4$) shells, depicted in \figref{fig1} and described in
detail in [SM]. The structures rest on a low-index
($\varepsilon\approx 1$) transparent substrate which not only provides
mechanical support but also a means of dissipating heat away from the
structure; the bottom of the substrate is assumed to be in contact
with a 300~K heat reservoir while surfaces exposed to vacuum satisfy
adiabatic boundary conditions ($\nabla T \cdot \vec{\hat{n}} =
0$). When heated by a highly conductive 2D material (e.g. carbon
nanotube wall or graphene sheet) at the GST--substrate interface, such
a structure can exhibit large temperature gradients within the core, a
consequence of boundary resistance between the various interfaces and
rapid heat dissipation in the highly conductive
shells~\cite{reifenberg2008impact,xiong2009inducing,liang2012ultra}. To
model the corresponding steady-state temperature distribution
$T(\vec{x})$, we solve the heat-conduction equation via
COMSOL~\footnote{Note that at these temperatures convective and
  radiative effects are negligible compared to conductive transfer,
  allowing us to consider the radiation and conduction problems
  separately.}, including the full temperature-dependent thermal
conductivity $\kappa(T)$ of the GST~\cite{lyeo2006thermal}. Note that
even at large temperatures, $\kappa_\mathrm{GST}(800~\mathrm{K}) \ll
\kappa_\mathrm{Ti,Si_3N_4}(300~\mathrm{K}) \gtrsim
20~W/\mathrm{m}\mathrm{K}$. The existence of (intrinsic and contact)
boundary resistance at this scale is taken into account by the
introduction of effective resistances $R_\mathrm{sh|c}$,
$R_\mathrm{h|su}$, and $R_\mathrm{sh|su}$, at the interfaces between
shell--GST, heater--substrate, and shell--substrate, respectively
[SM]. \Figref{fig2}(d) shows $T(\vec{x})$ throughout the Ti structure
when the GST--substrate interface is heated to $T_\mathrm{GST} =
870$~K (approaching the GST melting
temperature~\cite{tsafack2011electronic}), and under various operating
conditions. Specifically, we consider $R_{\mathrm{sh|su}}=10^{-8}
\mathrm{m}^2 W/$K and $R_{\mathrm{sh|c}}=R_{\mathrm{h|su}}=R_{th}$,
where \{(i), (ii), (iii)\} correspond to typical values of
$R_{th}=\{0.5, 1, 2\} \times 10^{-7} \mathrm{m}^2 W/$K while (iv)
$R_{th} = \infty$ and (v) $R_{th} = 0$ describe extreme situations
involving either perfect temperature localization in the GST or
uniform temperature throughout the structure, respectively.

%(i) $R_{th}=0.5\times 10^{-7} \mathrm{m}^2 W/$K, (ii) $R_{th}=10^{-7}
%\mathrm{m}^2 W/$K, (iii) $R_{th}=2\times 10^{-7} \mathrm{m}^2
%W/$K. Also considered are the extreme scenarios (iv) $R_{th} =
%\infty$ and (v) $R_{th} = 0$, correspondding to either perfect
%temperature localization in the GST or uniform temperature
%distributions throughout the structure. Higher temperature
%localization is possible by increasing the contact resistance between
%the shell and core.

%NOTE: Presence of T gradient => GST is in multi-phase with very
%different conductivities and permittivity across object. NOTE: This
%combination of materials also simultaneously satisfies good conditions
%for both large and directional emissivities.

\begin{figure*}[t!]
\begin{center}
\includegraphics[width=1\columnwidth]{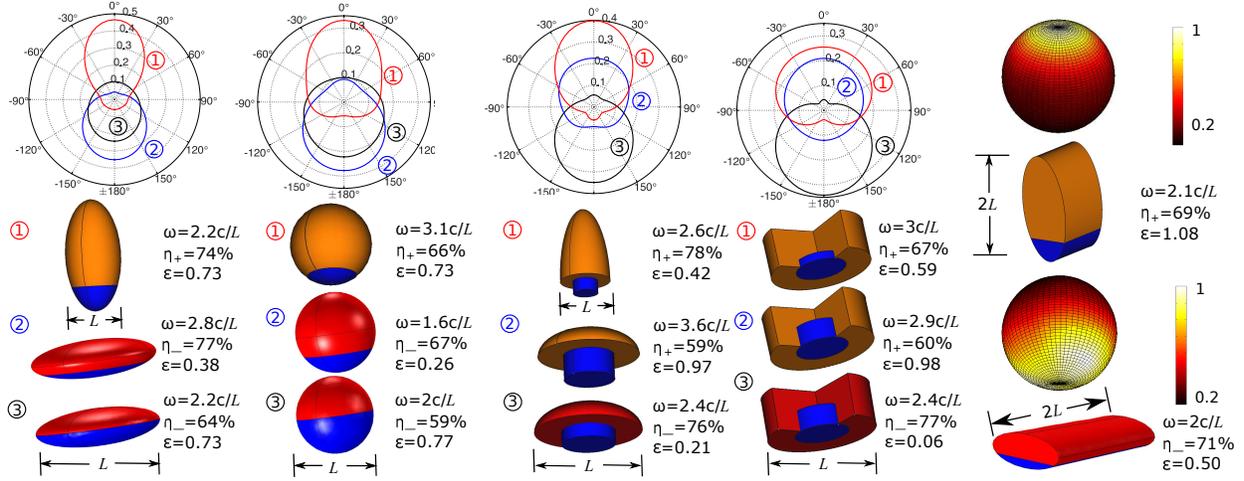}
\caption{Angular radiation intensity $U(\theta)$ for a variety of
  heterogeneous bodies---composite GST(blue), AZO (red), and
  Si$_3$N$_4$ (orange) ellipsoids, spheres, mushrooms and
  cylinders---at selected frequencies $\omega$. The temperature of the
  GST is held at 870~K while that of other materials is fixed at
  300~K. $\epsilon$ and $\eta_{\pm}$ denote the emissivity and partial
  directivities in the upper/lower hemispheres, defined (along with
  $\theta$) in \figref{fig1} and in the text. Polar (3D) plots are
  normalized by the total flux.}
\label{fig:fig3}
\end{center}
\end{figure*}

Given $T(\vec{x})$ and the dielectric properties of the
bodies~\cite{shportko2008resonant,li2008high,mash1973optical,kischkat2012mid},
the flux can be obtained via~\eqref{Uangle}. Due to temperature
gradients and phase transitions in the GST~\cite{lyeo2006thermal}, its
dielectric response $\varepsilon_\mathrm{GST}(T(\vec{x}),\omega)$
consists of continuously varying rather than piece-wise constant
regions [SM]; our FVC method, however, can handle arbitrarily varying
$\varepsilon(\vec{x})$ and $T(\vec{x})$. The choice of materials,
shapes, and dimensions of the hemispheroids ensure the existence of
geometric resonances near the thermal wavelength $\lambda_T \approx
5.8\mu$m, corresponding to the peak of the blackbody spectrum at
$870$~K. In this wavelength regime, $\varepsilon_\mathrm{GST}\approx
30+5i$~\cite{li2008high,shportko2008resonant},
$\varepsilon_\mathrm{Ti}\approx -100+80i$~\cite{mash1973optical}, and
$\varepsilon_{\mathrm{Si_3N_4}}\approx 5+0.1i$~\cite{kischkat2012mid},
enabling significant Purcell enhancement and
emission~\cite{bohren2008absorption}. \Figref{fig2} shows the
emissivity and partial directivity of the (a) Si$_3$N$_4$ and (b) Ti
structures, along with their corresponding $T(\vec{x})$ (insets) under
heating scenarios (ii) or (iii), respectively, and for $T_\mathrm{GST}
= 870$~K. The emissivity $\epsilon(\omega) =
\Phi(\omega)/\Phi_\mathrm{BB}(\omega)$ (blue dots) is defined as the
ratio of the thermal flux $\Phi(\omega)$ from each object to that of a
blackbody
$\Phi_{\mathrm{BB}}(\omega)=\frac{A}{4\pi^2}(\omega/c)^2\Theta(\omega,T)$
of the same surface area $A$ and uniform $T=870$~K (green lines); the
partial directivity $\eta_{\pm} = \Phi_{\pm} / \Phi$ (red lines) is
defined as the ratio of the flux into the upper/lower hemisphere
$\Phi_{\pm}(\omega) = 2\pi \int_{\mp \pi/2}^{\pi\mp\pi/2} d\theta \,
\Phi(\omega,\theta)$, to the total flux $\Phi$, where $\theta$ is
defined with respect to the $+\hat{z}$ axis [\figref{fig1}]. Note that
although $\epsilon$ exhibits multiple peaks, its magnitude ($\epsilon
\lesssim 0.2$) is limited by material losses ($\Im \varepsilon
\lesssim \Re \varepsilon$) in this frequency
range~\cite{bohren2008absorption}; larger $\epsilon$ can likely be
optained with further design and/or material combinations.

We find that $\eta$ increases sharply as the system transitions from
quasistatic to wavelength-scale behavior (in contrast to $\epsilon$
which exhibits gradual variations, except near a resonance). At small
$\omega L/c \ll 1$, the emission is highly quasi-isotropic (as
expected from a randomly polarized dipolar
emitter~\cite{greffet2002coherent}), becoming increasingly asymmetric
as $\omega L/c \gtrsim 1$. Essentially, with the help of the
curvature~\cite{yu2013enhancing}, the Si$_3$N$_4$ and Ti shells
redirect radiation upwards or downwards, enabling strong
\emph{coherent} interference between the radiated and scattered fields
of dipole emitters within the GST, making the design of the
temperature profile an essential ingredient for achieving large
$\eta$. \Figref{fig2}(c) shows the angular radiation intensity
$U(\theta)$ of the Si$_3$N$_4$ (red) and Ti (blue) structures at
selected frequencies and under two of the above-mentioned heating
conditions, corresponding to either (ii) partial temperature
localization in the GST (solid lines) or (v) uniform temperature
throughout the bodies (dashed lines). The dramatically different
radiation patterns and significantly smaller $\eta$ under (v) belie
the fact that dipole emitters inside the GST contribute larger partial
directivity compared to those in the shell, which tend to radiate
quasi-isotropically and dominate $\epsilon$.

To illustrate the non-negligible impact of $T(\vec{x})$ on the total
radiation of the bodies, \figref{fig2}(d) shows the
frequency-integrated, downward partial directivity $P_-=H_{-}/H$ of
the Ti structure under different $T_\mathrm{GST}$ and heating
scenarios, where $H_- = \int_0^\infty d\omega \, \Phi_-(\omega)$ and
$H=\int_0^\infty d\omega \, \Phi(\omega)$. As expected, $P_-$ grows
with increasing temperature localization in the GST, and remains
almost constant $\approx 0.55$ under uniform temperature
conditions. Such an increase in partial directivity, however, comes at
the expense of decreasing $\epsilon$ (not shown) due to the
increasingly dominant role of larger frequencies. At large $\omega$ or
for large bodies (where ray optics becomes valid), material losses
severely diminish $\epsilon$. Not surprisingly, the design criteria of
such wavelength-scale emitters differs significantly from that of
large-scale bodies (where Kirchoff's law is
valid~\cite{weinstein1960validity}). For instance, while larger $\eta$
can be obtained in the ray-optics limit by increasing the shell
thickness of each structure relative to the GST dimensions (thereby
enhancing extraction/reflections of radiation from the core), optimal
$\eta$ at a fixed frequency $\omega L/c \sim 1$ occur at specific
shell thicknesses, determined by the shape and materials of the
bodies. Planar structures can also yield highly directional emission
when subject to inhomogeneous temperature
distributions~\cite{wang2011direct}, but require significantly larger
boundary resistance (heating power) and offer limited degrees of
freedom for controlling emission. Compared to large-scale or planar
radiators, wavelength-scale composite bodies not only provide a high
degree of temperature tunability, but also enable simultaneous
enhancement in $\eta$ and $\epsilon$, even potentially exceeding the
ray-optical, blackbody limit~\cite{bohren2008absorption}.

\Figref{fig3} shows the radiation pattern of other heterogeneous
structures (ellipsoids, spheres, mushroom-like particles, and
cylindrical composites), depicted schematically in the figure with
blue/red/orange denoting GST/AZO/Si$_3$N$_4$ materials. Their shapes
and dimensions are detailed in [SM]. For simplicity, we consider
emission at selected frequencies and under heating scenario (iv),
corresponding to perfect temperature localization in the GST. As
expected, the design criteria for achieving large $\eta$ differs
depending on the choice of materials, with GST--Si$_3$N$_4$ composites
favoring large-curvature prolate bodies~\cite{yu2013enhancing} and
GST--AZO composites favoring oblate structures that provide higher
reflections.

\emph{Concluding remarks.---} The predictions above provide proof of
principle that combining conductive and radiative design principles in
wavelength-scale structures can lead to unusual thermal radiative
effects. Together with our FVC formalism, they motivate the need for
rigorous theoretical calculations of thermal emission that account for
existence of temperature and dielectric gradients in micron-scale,
structured surfaces, an issue that is especially relevant to thermal
metrology~\cite{fischer2005temperature}. The FVC framework not only
enables fast and accurate calculations, but also for techniques from
microwave antenna design and related fields to be carried over over to
problems involving infrared thermal radiation. Although the focus of
this work is on thermal radiation, similar ideas and techniques are
applicable to problems involving fluorescence or spontaneous emission
where, rather than controlling the temperature profile, it is possible
to localize and control the sources of emission via
doping~\cite{banaei2013design} or judicious choice of incident laser
light~\cite{LeRu08}.

We are grateful to Bhavin Shastri for very helpful comments. This work
was supported in part by the National Science Foundation under Grant
No.  DMR-1454836.

%%%%%%%%%% Merge with supplemental materials %%%%%%%%%%
\pagebreak
\begin{center}
  \textbf{\large Supplemental Materials: Temperature control of
    thermal radiation from heterogeneous bodies}
\end{center}

%%%%%%%%%% Prefix a "S" to all equations, figures, tables and reset the counter %%%%%%%%%%
\setcounter{equation}{0}
\setcounter{figure}{0}
\setcounter{table}{0}
\setcounter{page}{1}
\makeatletter
\renewcommand{\theequation}{S\arabic{equation}}
\renewcommand{\thefigure}{S\arabic{figure}}
%%%%%%%%%% Prefix a "S" to all equations, figures, tables and reset the counter %%%%%%%%%%
Below, we provide details of the geometric and material properties of
the composite bodies described in the main text, along with addition
discussion of the parameters and assumptions of the heating schemes
leading to temperature gradients.

\section{Geometry and material parameters}

\begin{figure}[h!]
\begin{center}
\includegraphics[width=0.5\columnwidth]{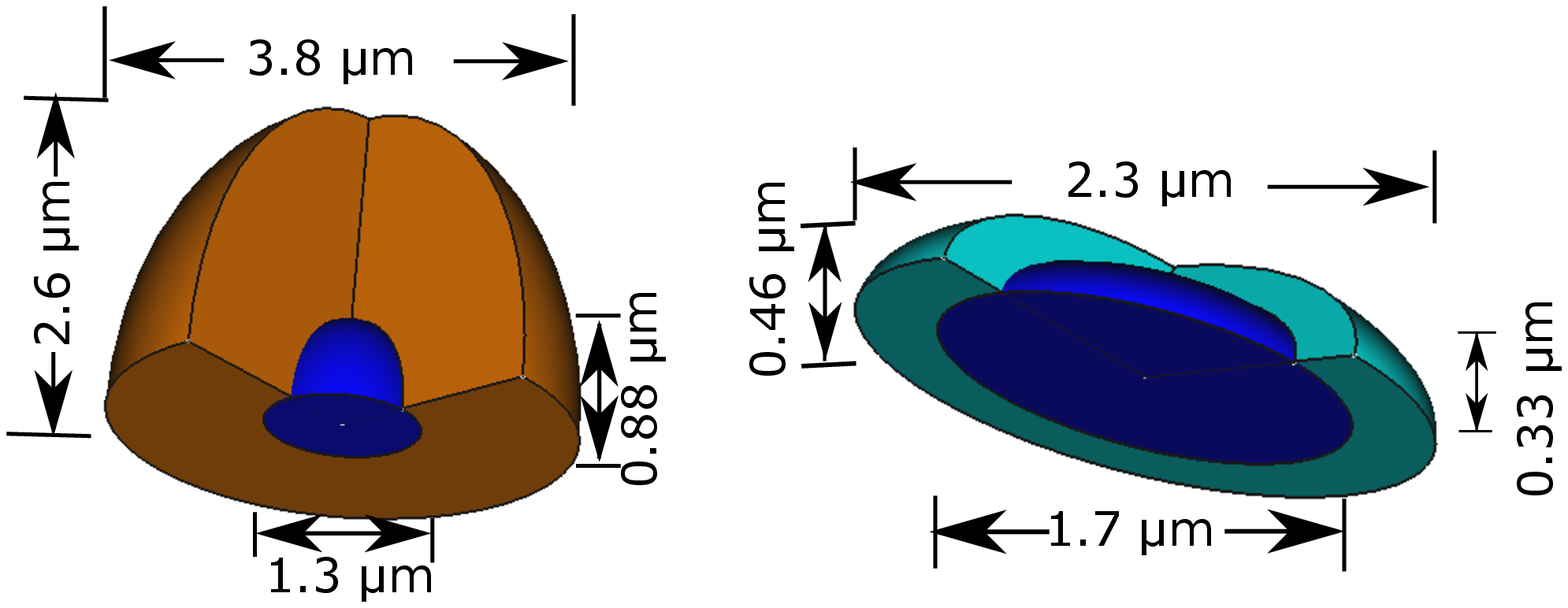}
\caption{Shape parameters describing the Si$_3$N$_4$ (left) and Ti
  (right) composite hemispheroids explored in Fig.~2 of the main
  text. Blue, orange, and green colors denote GST, Si$_3$N$_4$, Ti,
  respectively.}
\label{fig:sizehemi}
\end{center}
\end{figure}

\Figref{sizehemi} provides schematics of the Si$_3$N$_4$ (left) and Ti
(right) hemispheroid composites explored in Fig.~2 of the text, along
with the corresponding geometrical parameters. In particular, the
Si$_3$N$_4$ and Ti shells have long (short) semi-axes $2.6(1.9)\mu$m
and $1.2 (0.46)\mu$m, respectively; the chalcogenide (GST) cores have
long (short) semi-axes of $0.88 (0.63)\mu$m and $0.83 (0.33)\mu$m,
respectively. As noted in the text, these values are chosen in order
to enhance the emissivity and partial directivity of the structures.
\Figref{sizegeo} provides the size and dimensions of the various
geometries explored in Fig.~3 of the text.

%%%%%%%%%%%%%%%%%%%%%%%%%%%%%%
\begin{figure}[t!]
\begin{center}
\includegraphics[width=0.5\columnwidth]{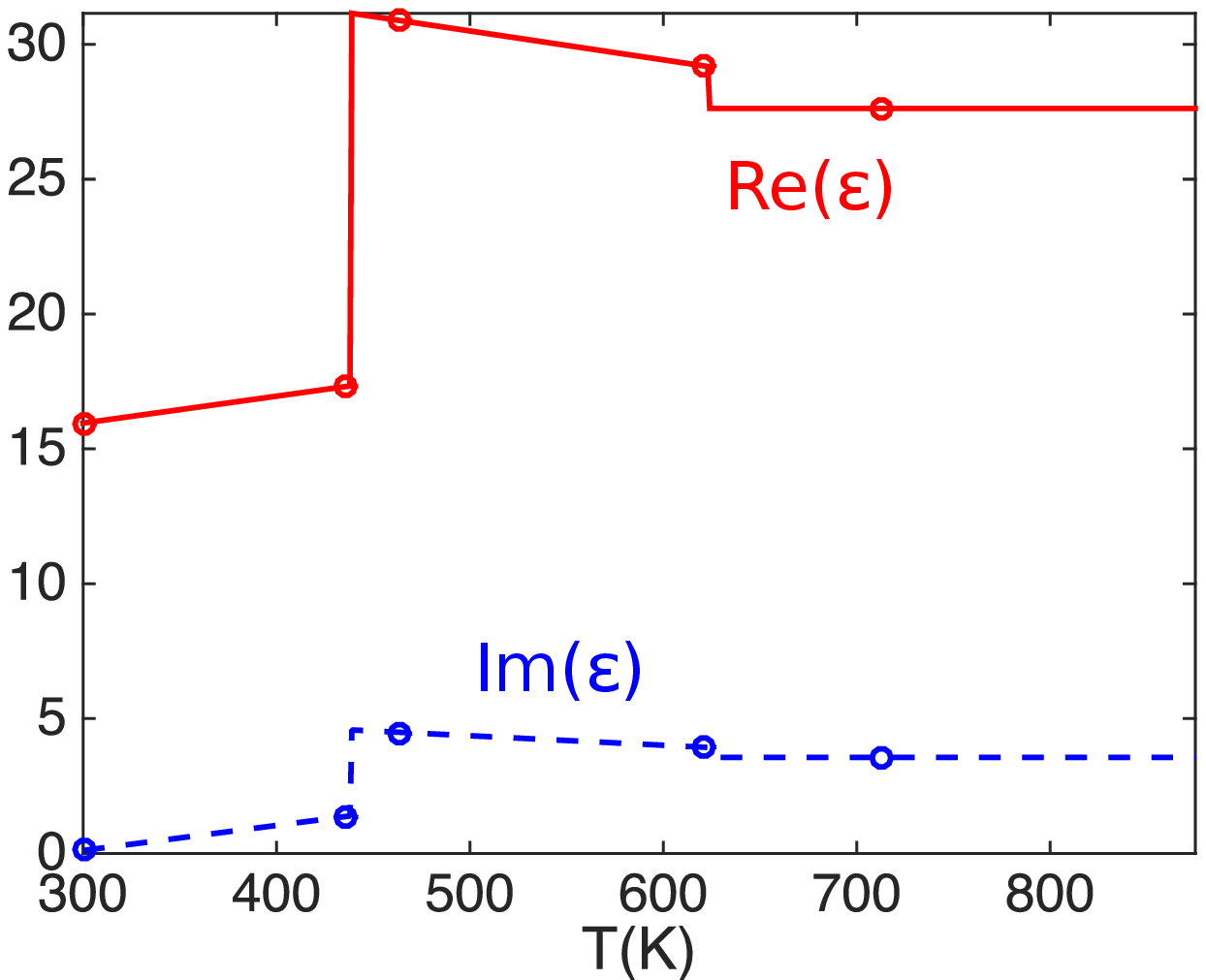}
\caption{Real (solid red) and imaginary (dashed blue) permittivity
  $\varepsilon_\mathrm{GST}(T,\lambda)$ of a bulk GST glass at
  $\lambda=5.8\mu$m, obtained via simple linear interpolation of
  experimental data at multiple temperatures (circles).}
\label{fig:disperse}
\end{center}
\end{figure}

The Ge$_2$Sb$_2$Te$_5$ alloy is a phase-change chalcogenide glass that
exhibits a large thermo-optic effect~\cite{rude2013optical} and three
possible (amorphous, cubic, and hexagonal) phases corresponding to
transition temperatures of 438~K (separating the amorphous and cubic
phases) and 623~K (separating the cubic and hexagonal
phases)~\cite{li2008high,xiong2009inducing}. Because there are yet no
experimental characterizations or semi-analytical models of the
dielectric dispersion $\varepsilon_\mathrm{GST}(\omega,T)$ of the GST
from $300$~K to its melting point
$870$~K~\cite{tsafack2011electronic}, we instead model the dispersion
via a simple linear-interpolated fit of available experimental data at
five different temperatures (spanning amorphous, cubic and hexagonal
phases)~\cite{li2008high, shportko2008resonant}. \Figref{disperse}
shows both the real (red solid line) and imaginary (blue dashed line)
parts of $\varepsilon_\mathrm{GST}$ at a single wavelength
$\lambda=5.8\mu$m over this temperature range (with circles denoting
experimental data). Together with the temperature profiles of the
structures $T(\vec{x})$ and dispersion relations of
Ti~\cite{mash1973optical}, Si$_3$N$_4$~\cite{kischkat2012mid}, and
AZO~\cite{kim2012plasmonic}, this provides all of the information
needed to perform the calculations of thermal radiation from the
bodies of Fig.~2 in the main text.  On the other hand, Fig.~ 3 of the
main text explores single-frequency radiation from bodies with
piece-wise constant temperature profiles (constant $T=870$~K in the
GST and $300$~K in the remaining regions), which allows us to employ
typical permittivity values for these materials at mid-infrared
wavelengths; specifically, we choose
$\varepsilon_{\mathrm{GST}}=30+10i$~\cite{shportko2008resonant},
$\varepsilon_{\mathrm{Ti}}=-100+80i$~\cite{mash1973optical},
$\varepsilon_{\mathrm{Si_3N_4}}=5+0.1i$~\cite{kischkat2012mid}, and
$\varepsilon_{\mathrm{AZO}}=-25+15i$, corresponding to a doping
density $\approx 1\mathrm{wt}\%$~\cite{kim2012plasmonic}.

%%%%%%%%%%

\section{Temperature gradients}

Interfaces play a crucial role in nanoscale thermal transport. For
instance, they enable thermal boundary resistance (TBR) to radically
alter the surrounding temperature
distribution~\cite{reifenberg2008impact,merabia2009critical,stevens2007effects,marconnet2013thermal},
leading to small-scale thermal discontinuities across the
interface. TBR consists of both contact and intrinsic ``Kapitza''
resistance, with the former arising from poor mechanical connection
between materials (due to surface roughness) and the latter from
acoustic mismatch between materials (and hence persisting even under
perfect-contact situations)~\cite{stevens2007effects}. Typical values
of intrinsic resistance at room temperature are on the order of
$10^{-9} \sim 10^{-7} \mathrm{m}^2 W/$K~\cite{stevens2007effects},
whereas those arising from contact resistance vary depending on the
surface and thermophysical properties of the intervening medium. In
our setup (described schematically in Fig.~1 of the main text), there
four interfaces at which TBR can arise. These are denoted and
described by the resistances $R_\mathrm{sh|c}$, $R_\mathrm{h|su}$,
$R_\mathrm{sh|su}$, and $R_{\mathrm{h|c}}$, of the shell--GST,
heater--substrate, shell--substrate, and heater--GST interfaces,
respectively. Note that the thermal resistance associated with
graphene can be made extremely small~\cite{shahil2012thermal} and
hence in our calculations, we assume negligible $R_{\mathrm{h|c}} =
0$. In order to obtain large temperature gradients, it is important to
operate with materials that can dissipate heat away from the shells
rapidly~\cite{islam2013role}; hence, we assume small shell--substrate
interface resistances $R_{\mathrm{sh|su}}=10^{-8} \mathrm{m}^2
W/$K. For simplicity, we consider conditions under which the interface
resistances $R_{\mathrm{sh|c}}=R_{\mathrm{h|su}}=R_\mathrm{th}$ are
equal and obtain various degrees of temperature localization by
varying $R_{th}$, with $R_{th}=\infty$ leading to perfect
temperature-localization and $R_{th}=0$ leading to uniform temperature
distributions. In particular, we consider five different operating
conditions, corresponding to realistic values of (i) $R_{th}=0.5\times
10^{-7} \mathrm{m}^2 W/$K, (ii) $R_{th}=10^{-7} \mathrm{m}^2 W/$K, and
(iii) $R_{th}=2\times 10^{-7} \mathrm{m}^2 W/$K and extreme
(unrealistic) values of (iv) $R_{th} = \infty$, and (v) $R_{th} = 0$.

\begin{figure}[t!]
\begin{center}
\includegraphics[width=0.5\columnwidth]{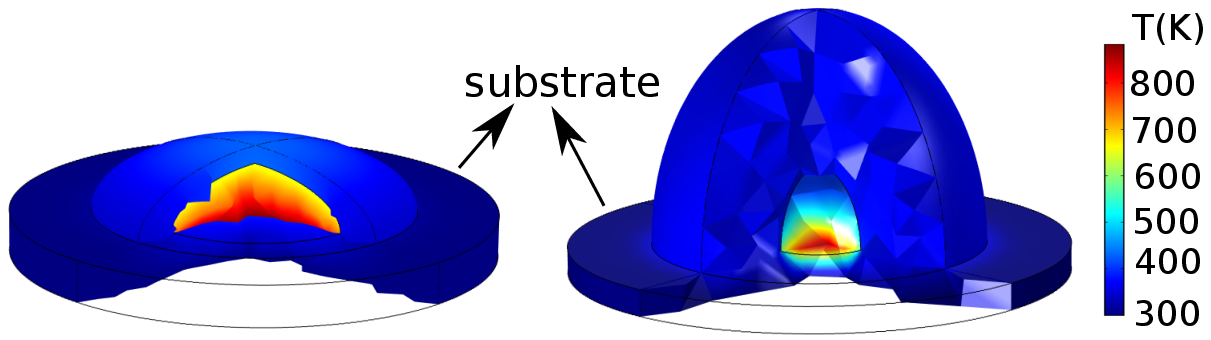}
\caption{Temperature distribution $T(\vec{x})$ of the Ti (left) and
  Si$_3$N$_4$ (right) hemispheroid composites described in
  Fig.~1. Both structures rest on a SiO$_2$ substrate (thickness
  0.3$\mu$m and radius $=1.5\times$ shell radius) whose bottom surface
  is in contact with a heat reservoir at 300~K. All other surfaces are
  exposed to vacuum and therefore satisfy adiabatic boundary
  conditions; material interfaces on the other hand are subject to
  thermal boundary resistance in accordance with operating condition
  (iii) and (ii) described in the text, for the left and right body,
  respectively.}
\label{fig:tem}
\end{center}
\end{figure}

\begin{table}[t!]
\renewcommand{\arraystretch}{1.0}
\caption{The thermal properties used in our COMSOL simulation} \label{tab:parameter} \centering
\footnotesize
\begin{tabular}{c | c | c}
  \hline\hline
               &k(W/m/K)    &C(J/kg/K)   \\[0.4em] \hline
     Ti        &21.9        &523      \\[0.3em]\hline
     Si$_3$N$_4$ &40          &1100     \\[0.3em]\hline
     GST       &$\kappa(T)$~\cite{lyeo2006thermal} &208.3  \\[0.3em]\hline
     SiO$_2$   &1.38        &703      \\[0.3em]\hline\hline
\end{tabular}
\end{table}

In order to solve the heat--conduction equation to obtain the
steady-state temperature distribution $T(\vec{x})$, one must also
specify the boundary conditions associated with vacuum--material
interfaces, which we assume to be adiabatic ($\nabla T \cdot
\vec{\hat{n}} = 0$), corresponding to negligible conduction,
convection, and radiative-heat dissipation through air. The substrate
is chosen to be a 0.3$~\mu$m thick SiO$_2$ film in contact with a heat
reservoir at 300~K through the bottom interface, leading to large heat
dissipation away from the shell and hence large temperature
localization in the GST with decreasing substrate thickness. We choose
the substrate lateral (cylindrical) dimensions to be large enough to
remove large thermal diffusion away from the GST.  \Figref{tem} shows
the temperature distribution of both Ti (left) and Si$_3$N$_4$ (right)
structures under the operating condition (iii) and (ii), respectively,
assuming the material conductivities and heat capacities given in
Table I. As shown, the temperature in the substrate and shells is
almost uniform and close to 300~K, thanks to the presence of boundary
resistance between the heater and substrate (which bars heat from
flowing into the substrate) along with the high thermal conductivities
of Ti and Si$_3$N$_4$, which act to dissipate heat away from the GST.

\begin{figure*}[t!]
\begin{center}
\includegraphics[width=\columnwidth]{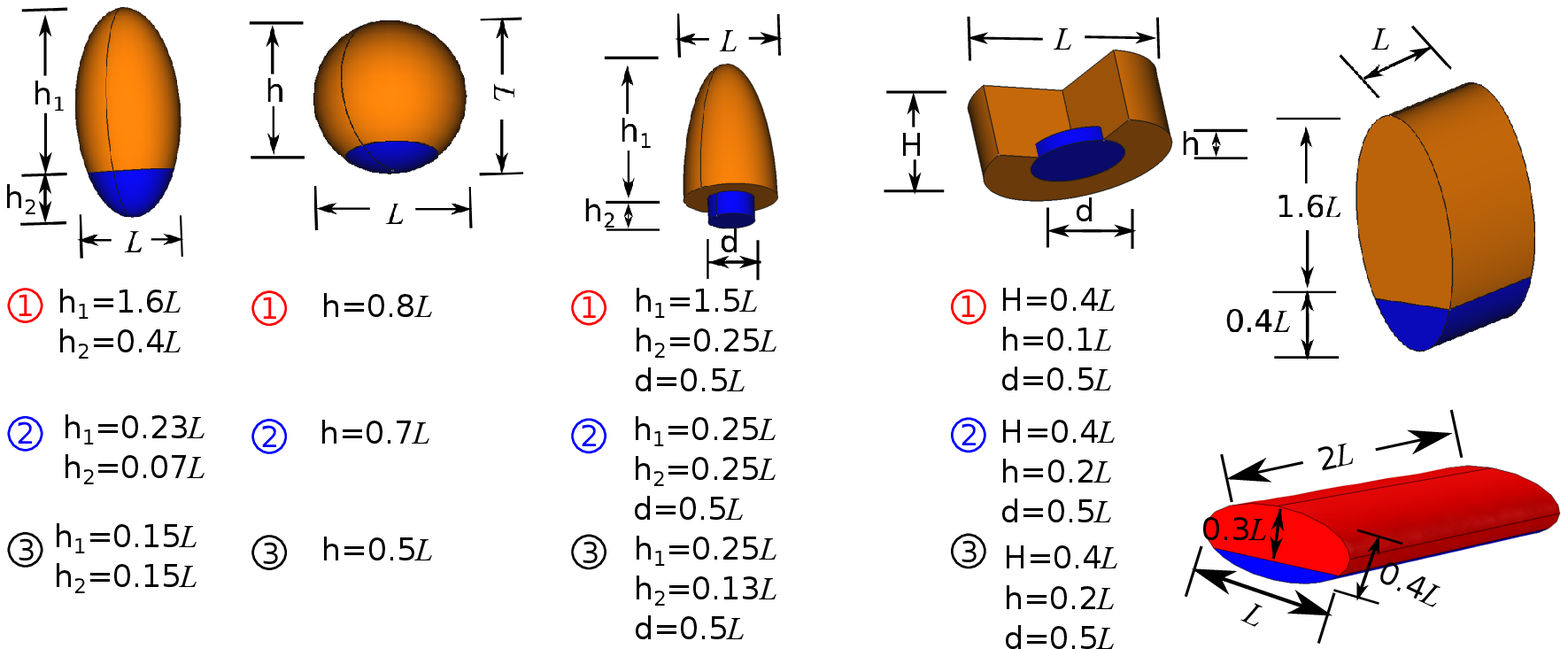}
\caption{Parameter descriptions for the bodies associated with Fig.~3
  of the main text.}
\label{fig:sizegeo}
\end{center}
\end{figure*}

\bibliographystyle{apsrev}
\bibliography{ref,photon}

\begin{thebibliography}{15}
\expandafter\ifx\csname natexlab\endcsname\relax\def\natexlab#1{#1}\fi
\expandafter\ifx\csname bibnamefont\endcsname\relax
  \def\bibnamefont#1{#1}\fi
\expandafter\ifx\csname bibfnamefont\endcsname\relax
  \def\bibfnamefont#1{#1}\fi
\expandafter\ifx\csname citenamefont\endcsname\relax
  \def\citenamefont#1{#1}\fi
\expandafter\ifx\csname url\endcsname\relax
  \def\url#1{\texttt{#1}}\fi
\expandafter\ifx\csname urlprefix\endcsname\relax\def\urlprefix{URL }\fi
\providecommand{\bibinfo}[2]{#2}
\providecommand{\eprint}[2][]{\url{#2}}

\bibitem[{\citenamefont{Rud{\'e} et~al.}(2013)\citenamefont{Rud{\'e}, Pello,
  Simpson, Osmond, Roelkens, van~der Tol, and Pruneri}}]{rude2013optical}
\bibinfo{author}{\bibfnamefont{M.}~\bibnamefont{Rud{\'e}}},
  \bibinfo{author}{\bibfnamefont{J.}~\bibnamefont{Pello}},
  \bibinfo{author}{\bibfnamefont{R.~E.} \bibnamefont{Simpson}},
  \bibinfo{author}{\bibfnamefont{J.}~\bibnamefont{Osmond}},
  \bibinfo{author}{\bibfnamefont{G.}~\bibnamefont{Roelkens}},
  \bibinfo{author}{\bibfnamefont{J.~J.} \bibnamefont{van~der Tol}},
  \bibnamefont{and} \bibinfo{author}{\bibfnamefont{V.}~\bibnamefont{Pruneri}},
  \bibinfo{journal}{Applied Physics Letters} \textbf{\bibinfo{volume}{103}},
  \bibinfo{pages}{141119} (\bibinfo{year}{2013}).

\bibitem[{\citenamefont{Li et~al.}(2008)\citenamefont{Li, Choi, Byun, Kim, Sim,
  and Kim}}]{li2008high}
\bibinfo{author}{\bibfnamefont{X.~Z.} \bibnamefont{Li}},
  \bibinfo{author}{\bibfnamefont{J.~K.} \bibnamefont{Choi}},
  \bibinfo{author}{\bibfnamefont{Y.~S.} \bibnamefont{Byun}},
  \bibinfo{author}{\bibfnamefont{S.~Y.} \bibnamefont{Kim}},
  \bibinfo{author}{\bibfnamefont{K.~S.} \bibnamefont{Sim}}, \bibnamefont{and}
  \bibinfo{author}{\bibfnamefont{S.~K.} \bibnamefont{Kim}},
  \bibinfo{journal}{Japanese Journal of Applied Physics}
  \textbf{\bibinfo{volume}{47}}, \bibinfo{pages}{5477} (\bibinfo{year}{2008}).

\bibitem[{\citenamefont{Xiong et~al.}(2009)\citenamefont{Xiong, Liao, and
  Pop}}]{xiong2009inducing}
\bibinfo{author}{\bibfnamefont{F.}~\bibnamefont{Xiong}},
  \bibinfo{author}{\bibfnamefont{A.}~\bibnamefont{Liao}}, \bibnamefont{and}
  \bibinfo{author}{\bibfnamefont{E.}~\bibnamefont{Pop}},
  \bibinfo{journal}{Applied Physics Letters} \textbf{\bibinfo{volume}{95}},
  \bibinfo{pages}{243103} (\bibinfo{year}{2009}).

\bibitem[{\citenamefont{Tsafack et~al.}(2011)\citenamefont{Tsafack, Piccinini,
  Lee, Pop, and Rudan}}]{tsafack2011electronic}
\bibinfo{author}{\bibfnamefont{T.}~\bibnamefont{Tsafack}},
  \bibinfo{author}{\bibfnamefont{E.}~\bibnamefont{Piccinini}},
  \bibinfo{author}{\bibfnamefont{B.-S.} \bibnamefont{Lee}},
  \bibinfo{author}{\bibfnamefont{E.}~\bibnamefont{Pop}}, \bibnamefont{and}
  \bibinfo{author}{\bibfnamefont{M.}~\bibnamefont{Rudan}},
  \bibinfo{journal}{Journal of Applied Physics} \textbf{\bibinfo{volume}{110}},
  \bibinfo{pages}{063716} (\bibinfo{year}{2011}).

\bibitem[{\citenamefont{Shportko et~al.}(2008)\citenamefont{Shportko, Kremers,
  Woda, Lencer, Robertson, and Wuttig}}]{shportko2008resonant}
\bibinfo{author}{\bibfnamefont{K.}~\bibnamefont{Shportko}},
  \bibinfo{author}{\bibfnamefont{S.}~\bibnamefont{Kremers}},
  \bibinfo{author}{\bibfnamefont{M.}~\bibnamefont{Woda}},
  \bibinfo{author}{\bibfnamefont{D.}~\bibnamefont{Lencer}},
  \bibinfo{author}{\bibfnamefont{J.}~\bibnamefont{Robertson}},
  \bibnamefont{and} \bibinfo{author}{\bibfnamefont{M.}~\bibnamefont{Wuttig}},
  \bibinfo{journal}{Nature materials} \textbf{\bibinfo{volume}{7}},
  \bibinfo{pages}{653} (\bibinfo{year}{2008}).

\bibitem[{\citenamefont{Mash and Motulevich}(1973)}]{mash1973optical}
\bibinfo{author}{\bibfnamefont{I.}~\bibnamefont{Mash}} \bibnamefont{and}
  \bibinfo{author}{\bibfnamefont{G.}~\bibnamefont{Motulevich}},
  \bibinfo{journal}{SOVIET PHYSICS JETP} \textbf{\bibinfo{volume}{36}}
  (\bibinfo{year}{1973}).

\bibitem[{\citenamefont{Kischkat et~al.}(2012)\citenamefont{Kischkat, Peters,
  Gruska, Semtsiv, Chashnikova, Klinkm{\"u}ller, Fedosenko, Machulik,
  Aleksandrova, Monastyrskyi et~al.}}]{kischkat2012mid}
\bibinfo{author}{\bibfnamefont{J.}~\bibnamefont{Kischkat}},
  \bibinfo{author}{\bibfnamefont{S.}~\bibnamefont{Peters}},
  \bibinfo{author}{\bibfnamefont{B.}~\bibnamefont{Gruska}},
  \bibinfo{author}{\bibfnamefont{M.}~\bibnamefont{Semtsiv}},
  \bibinfo{author}{\bibfnamefont{M.}~\bibnamefont{Chashnikova}},
  \bibinfo{author}{\bibfnamefont{M.}~\bibnamefont{Klinkm{\"u}ller}},
  \bibinfo{author}{\bibfnamefont{O.}~\bibnamefont{Fedosenko}},
  \bibinfo{author}{\bibfnamefont{S.}~\bibnamefont{Machulik}},
  \bibinfo{author}{\bibfnamefont{A.}~\bibnamefont{Aleksandrova}},
  \bibinfo{author}{\bibfnamefont{G.}~\bibnamefont{Monastyrskyi}},
  \bibnamefont{et~al.}, \bibinfo{journal}{Applied optics}
  \textbf{\bibinfo{volume}{51}}, \bibinfo{pages}{6789} (\bibinfo{year}{2012}).

\bibitem[{\citenamefont{Kim et~al.}(2012)\citenamefont{Kim, Naik, Emani, and
  Boltasseva}}]{kim2012plasmonic}
\bibinfo{author}{\bibfnamefont{J.}~\bibnamefont{Kim}},
  \bibinfo{author}{\bibfnamefont{G.~V.} \bibnamefont{Naik}},
  \bibinfo{author}{\bibfnamefont{N.~K.} \bibnamefont{Emani}}, \bibnamefont{and}
  \bibinfo{author}{\bibfnamefont{A.}~\bibnamefont{Boltasseva}},
  \bibinfo{journal}{arXiv preprint arXiv:1211.5988}  (\bibinfo{year}{2012}).

\bibitem[{\citenamefont{Reifenberg et~al.}(2008)\citenamefont{Reifenberg,
  Kencke, and Goodson}}]{reifenberg2008impact}
\bibinfo{author}{\bibfnamefont{J.~P.} \bibnamefont{Reifenberg}},
  \bibinfo{author}{\bibfnamefont{D.~L.} \bibnamefont{Kencke}},
  \bibnamefont{and} \bibinfo{author}{\bibfnamefont{K.~E.}
  \bibnamefont{Goodson}}, \bibinfo{journal}{Electron Device Letters, IEEE}
  \textbf{\bibinfo{volume}{29}}, \bibinfo{pages}{1112} (\bibinfo{year}{2008}).

\bibitem[{\citenamefont{Merabia et~al.}(2009)\citenamefont{Merabia, Keblinski,
  Joly, Lewis, and Barrat}}]{merabia2009critical}
\bibinfo{author}{\bibfnamefont{S.}~\bibnamefont{Merabia}},
  \bibinfo{author}{\bibfnamefont{P.}~\bibnamefont{Keblinski}},
  \bibinfo{author}{\bibfnamefont{L.}~\bibnamefont{Joly}},
  \bibinfo{author}{\bibfnamefont{L.~J.} \bibnamefont{Lewis}}, \bibnamefont{and}
  \bibinfo{author}{\bibfnamefont{J.-L.} \bibnamefont{Barrat}},
  \bibinfo{journal}{PRE} \textbf{\bibinfo{volume}{79}}, \bibinfo{pages}{021404}
  (\bibinfo{year}{2009}).

\bibitem[{\citenamefont{Stevens et~al.}(2007)\citenamefont{Stevens, Zhigilei,
  and Norris}}]{stevens2007effects}
\bibinfo{author}{\bibfnamefont{R.~J.} \bibnamefont{Stevens}},
  \bibinfo{author}{\bibfnamefont{L.~V.} \bibnamefont{Zhigilei}},
  \bibnamefont{and} \bibinfo{author}{\bibfnamefont{P.~M.}
  \bibnamefont{Norris}}, \bibinfo{journal}{International Journal of Heat and
  Mass Transfer} \textbf{\bibinfo{volume}{50}}, \bibinfo{pages}{3977}
  (\bibinfo{year}{2007}).

\bibitem[{\citenamefont{Marconnet et~al.}(2013)\citenamefont{Marconnet, Panzer,
  and Goodson}}]{marconnet2013thermal}
\bibinfo{author}{\bibfnamefont{A.~M.} \bibnamefont{Marconnet}},
  \bibinfo{author}{\bibfnamefont{M.~A.} \bibnamefont{Panzer}},
  \bibnamefont{and} \bibinfo{author}{\bibfnamefont{K.~E.}
  \bibnamefont{Goodson}}, \bibinfo{journal}{Reviews of Modern Physics}
  \textbf{\bibinfo{volume}{85}}, \bibinfo{pages}{1295} (\bibinfo{year}{2013}).

\bibitem[{\citenamefont{Shahil and Balandin}(2012)}]{shahil2012thermal}
\bibinfo{author}{\bibfnamefont{K.~M.} \bibnamefont{Shahil}} \bibnamefont{and}
  \bibinfo{author}{\bibfnamefont{A.~A.} \bibnamefont{Balandin}},
  \bibinfo{journal}{Solid State Communications} \textbf{\bibinfo{volume}{152}},
  \bibinfo{pages}{1331} (\bibinfo{year}{2012}).

\bibitem[{\citenamefont{Islam et~al.}(2013)\citenamefont{Islam, Li, Dorgan,
  Bae, and Pop}}]{islam2013role}
\bibinfo{author}{\bibfnamefont{S.}~\bibnamefont{Islam}},
  \bibinfo{author}{\bibfnamefont{Z.}~\bibnamefont{Li}},
  \bibinfo{author}{\bibfnamefont{V.~E.} \bibnamefont{Dorgan}},
  \bibinfo{author}{\bibfnamefont{M.-H.} \bibnamefont{Bae}}, \bibnamefont{and}
  \bibinfo{author}{\bibfnamefont{E.}~\bibnamefont{Pop}},
  \bibinfo{journal}{Electron Device Letters, IEEE}
  \textbf{\bibinfo{volume}{34}}, \bibinfo{pages}{166} (\bibinfo{year}{2013}).

\bibitem[{\citenamefont{Lyeo et~al.}(2006)\citenamefont{Lyeo, Cahill, Lee,
  Abelson, Kwon, Kim, Bishop, and Cheong}}]{lyeo2006thermal}
\bibinfo{author}{\bibfnamefont{H.-K.} \bibnamefont{Lyeo}},
  \bibinfo{author}{\bibfnamefont{D.~G.} \bibnamefont{Cahill}},
  \bibinfo{author}{\bibfnamefont{B.-S.} \bibnamefont{Lee}},
  \bibinfo{author}{\bibfnamefont{J.~R.} \bibnamefont{Abelson}},
  \bibinfo{author}{\bibfnamefont{M.-H.} \bibnamefont{Kwon}},
  \bibinfo{author}{\bibfnamefont{K.-B.} \bibnamefont{Kim}},
  \bibinfo{author}{\bibfnamefont{S.~G.} \bibnamefont{Bishop}},
  \bibnamefont{and} \bibinfo{author}{\bibfnamefont{B.-k.}
  \bibnamefont{Cheong}}, \bibinfo{journal}{Applied Physics Letters}
  \textbf{\bibinfo{volume}{89}}, \bibinfo{pages}{151904}
  (\bibinfo{year}{2006}).

\end{thebibliography}


\begin{thebibliography}{75}
\expandafter\ifx\csname natexlab\endcsname\relax\def\natexlab#1{#1}\fi
\expandafter\ifx\csname bibnamefont\endcsname\relax
  \def\bibnamefont#1{#1}\fi
\expandafter\ifx\csname bibfnamefont\endcsname\relax
  \def\bibfnamefont#1{#1}\fi
\expandafter\ifx\csname citenamefont\endcsname\relax
  \def\citenamefont#1{#1}\fi
\expandafter\ifx\csname url\endcsname\relax
  \def\url#1{\texttt{#1}}\fi
\expandafter\ifx\csname urlprefix\endcsname\relax\def\urlprefix{URL }\fi
\providecommand{\bibinfo}[2]{#2}
\providecommand{\eprint}[2][]{\url{#2}}

\bibitem[{\citenamefont{Greffet and Henkel}(2007)}]{greffet2007coherent}
\bibinfo{author}{\bibfnamefont{J.-J.} \bibnamefont{Greffet}} \bibnamefont{and}
  \bibinfo{author}{\bibfnamefont{C.}~\bibnamefont{Henkel}},
  \bibinfo{journal}{Contemporary Physics} \textbf{\bibinfo{volume}{48}},
  \bibinfo{pages}{183} (\bibinfo{year}{2007}).

\bibitem[{\citenamefont{Masuda et~al.}(1988)\citenamefont{Masuda, Takashima,
  and Takayama}}]{masuda1988emissivity}
\bibinfo{author}{\bibfnamefont{K.}~\bibnamefont{Masuda}},
  \bibinfo{author}{\bibfnamefont{T.}~\bibnamefont{Takashima}},
  \bibnamefont{and} \bibinfo{author}{\bibfnamefont{Y.}~\bibnamefont{Takayama}},
  \bibinfo{journal}{Remote Sensing of Environment}
  \textbf{\bibinfo{volume}{24}}, \bibinfo{pages}{313} (\bibinfo{year}{1988}).

\bibitem[{\citenamefont{Ilic and Solja{\v{c}}i{\'c}}(2014)}]{ilic2014thermal}
\bibinfo{author}{\bibfnamefont{O.}~\bibnamefont{Ilic}} \bibnamefont{and}
  \bibinfo{author}{\bibfnamefont{M.}~\bibnamefont{Solja{\v{c}}i{\'c}}},
  \bibinfo{journal}{Nature materials} \textbf{\bibinfo{volume}{13}},
  \bibinfo{pages}{920} (\bibinfo{year}{2014}).

\bibitem[{\citenamefont{Rinnerbauer et~al.}(2014)\citenamefont{Rinnerbauer,
  Lenert, Bierman, Yeng, Chan, Geil, Senkevich, Joannopoulos, Wang,
  Solja{\v{c}}i{\'c} et~al.}}]{rinnerbauer2014metallic}
\bibinfo{author}{\bibfnamefont{V.}~\bibnamefont{Rinnerbauer}},
  \bibinfo{author}{\bibfnamefont{A.}~\bibnamefont{Lenert}},
  \bibinfo{author}{\bibfnamefont{D.~M.} \bibnamefont{Bierman}},
  \bibinfo{author}{\bibfnamefont{Y.~X.} \bibnamefont{Yeng}},
  \bibinfo{author}{\bibfnamefont{W.~R.} \bibnamefont{Chan}},
  \bibinfo{author}{\bibfnamefont{R.~D.} \bibnamefont{Geil}},
  \bibinfo{author}{\bibfnamefont{J.~J.} \bibnamefont{Senkevich}},
  \bibinfo{author}{\bibfnamefont{J.~D.} \bibnamefont{Joannopoulos}},
  \bibinfo{author}{\bibfnamefont{E.~N.} \bibnamefont{Wang}},
  \bibinfo{author}{\bibfnamefont{M.}~\bibnamefont{Solja{\v{c}}i{\'c}}},
  \bibnamefont{et~al.}, \bibinfo{journal}{Advanced Energy Materials}
  \textbf{\bibinfo{volume}{4}} (\bibinfo{year}{2014}).

\bibitem[{\citenamefont{Fan}(2014)}]{fan2014photovoltaics}
\bibinfo{author}{\bibfnamefont{S.}~\bibnamefont{Fan}}, \bibinfo{journal}{Nature
  nanotechnology} \textbf{\bibinfo{volume}{9}}, \bibinfo{pages}{92}
  (\bibinfo{year}{2014}).

\bibitem[{\citenamefont{Bermel et~al.}(2011)\citenamefont{Bermel, Ghebrebrhan,
  Harradon, Yeng, Celanovic, Joannopoulos, and Soljacic}}]{bermel2011tailoring}
\bibinfo{author}{\bibfnamefont{P.}~\bibnamefont{Bermel}},
  \bibinfo{author}{\bibfnamefont{M.}~\bibnamefont{Ghebrebrhan}},
  \bibinfo{author}{\bibfnamefont{M.}~\bibnamefont{Harradon}},
  \bibinfo{author}{\bibfnamefont{Y.~X.} \bibnamefont{Yeng}},
  \bibinfo{author}{\bibfnamefont{I.}~\bibnamefont{Celanovic}},
  \bibinfo{author}{\bibfnamefont{J.~D.} \bibnamefont{Joannopoulos}},
  \bibnamefont{and} \bibinfo{author}{\bibfnamefont{M.}~\bibnamefont{Soljacic}},
  \bibinfo{journal}{Nanoscale research letters} \textbf{\bibinfo{volume}{6}},
  \bibinfo{pages}{1} (\bibinfo{year}{2011}).

\bibitem[{\citenamefont{Florescu et~al.}(2007)\citenamefont{Florescu, Lee,
  Puscasu, Pralle, Florescu, Ting, and Dowling}}]{florescu2007improving}
\bibinfo{author}{\bibfnamefont{M.}~\bibnamefont{Florescu}},
  \bibinfo{author}{\bibfnamefont{H.}~\bibnamefont{Lee}},
  \bibinfo{author}{\bibfnamefont{I.}~\bibnamefont{Puscasu}},
  \bibinfo{author}{\bibfnamefont{M.}~\bibnamefont{Pralle}},
  \bibinfo{author}{\bibfnamefont{L.}~\bibnamefont{Florescu}},
  \bibinfo{author}{\bibfnamefont{D.~Z.} \bibnamefont{Ting}}, \bibnamefont{and}
  \bibinfo{author}{\bibfnamefont{J.~P.} \bibnamefont{Dowling}},
  \bibinfo{journal}{Solar Energy Materials and Solar Cells}
  \textbf{\bibinfo{volume}{91}}, \bibinfo{pages}{1599} (\bibinfo{year}{2007}).

\bibitem[{\citenamefont{Cahill et~al.}(2003)\citenamefont{Cahill, Ford,
  Goodson, Mahan, Majumdar, Maris, Merlin, and Phillpot}}]{cahill2003nanoscale}
\bibinfo{author}{\bibfnamefont{D.~G.} \bibnamefont{Cahill}},
  \bibinfo{author}{\bibfnamefont{W.~K.} \bibnamefont{Ford}},
  \bibinfo{author}{\bibfnamefont{K.~E.} \bibnamefont{Goodson}},
  \bibinfo{author}{\bibfnamefont{G.~D.} \bibnamefont{Mahan}},
  \bibinfo{author}{\bibfnamefont{A.}~\bibnamefont{Majumdar}},
  \bibinfo{author}{\bibfnamefont{H.~J.} \bibnamefont{Maris}},
  \bibinfo{author}{\bibfnamefont{R.}~\bibnamefont{Merlin}}, \bibnamefont{and}
  \bibinfo{author}{\bibfnamefont{S.~R.} \bibnamefont{Phillpot}},
  \bibinfo{journal}{Journal of Applied Physics} \textbf{\bibinfo{volume}{93}},
  \bibinfo{pages}{793} (\bibinfo{year}{2003}).

\bibitem[{\citenamefont{Cahill et~al.}(2014)\citenamefont{Cahill, Braun, Chen,
  Clarke, Fan, Goodson, Keblinski, King, Mahan, Majumdar
  et~al.}}]{cahill2014nanoscale}
\bibinfo{author}{\bibfnamefont{D.~G.} \bibnamefont{Cahill}},
  \bibinfo{author}{\bibfnamefont{P.~V.} \bibnamefont{Braun}},
  \bibinfo{author}{\bibfnamefont{G.}~\bibnamefont{Chen}},
  \bibinfo{author}{\bibfnamefont{D.~R.} \bibnamefont{Clarke}},
  \bibinfo{author}{\bibfnamefont{S.}~\bibnamefont{Fan}},
  \bibinfo{author}{\bibfnamefont{K.~E.} \bibnamefont{Goodson}},
  \bibinfo{author}{\bibfnamefont{P.}~\bibnamefont{Keblinski}},
  \bibinfo{author}{\bibfnamefont{W.~P.} \bibnamefont{King}},
  \bibinfo{author}{\bibfnamefont{G.~D.} \bibnamefont{Mahan}},
  \bibinfo{author}{\bibfnamefont{A.}~\bibnamefont{Majumdar}},
  \bibnamefont{et~al.}, \bibinfo{journal}{Applied Physics Reviews}
  \textbf{\bibinfo{volume}{1}}, \bibinfo{pages}{011305} (\bibinfo{year}{2014}).

\bibitem[{\citenamefont{De~Zoysa et~al.}(2012)\citenamefont{De~Zoysa, Asano,
  Mochizuki, Oskooi, Inoue, and Noda}}]{de2012conversion}
\bibinfo{author}{\bibfnamefont{M.}~\bibnamefont{De~Zoysa}},
  \bibinfo{author}{\bibfnamefont{T.}~\bibnamefont{Asano}},
  \bibinfo{author}{\bibfnamefont{K.}~\bibnamefont{Mochizuki}},
  \bibinfo{author}{\bibfnamefont{A.}~\bibnamefont{Oskooi}},
  \bibinfo{author}{\bibfnamefont{T.}~\bibnamefont{Inoue}}, \bibnamefont{and}
  \bibinfo{author}{\bibfnamefont{S.}~\bibnamefont{Noda}},
  \bibinfo{journal}{Nature Photonics} \textbf{\bibinfo{volume}{6}},
  \bibinfo{pages}{535} (\bibinfo{year}{2012}).

\bibitem[{\citenamefont{Wang et~al.}(2014)\citenamefont{Wang, Fu, and
  Tan}}]{wang2014thermal}
\bibinfo{author}{\bibfnamefont{W.}~\bibnamefont{Wang}},
  \bibinfo{author}{\bibfnamefont{C.}~\bibnamefont{Fu}}, \bibnamefont{and}
  \bibinfo{author}{\bibfnamefont{W.}~\bibnamefont{Tan}},
  \bibinfo{journal}{Journal of Quantitative Spectroscopy and Radiative
  Transfer} \textbf{\bibinfo{volume}{132}}, \bibinfo{pages}{36}
  (\bibinfo{year}{2014}).

\bibitem[{\citenamefont{Greffet et~al.}(2002)\citenamefont{Greffet, Carminati,
  Joulain, Mulet, Mainguy, and Chen}}]{greffet2002coherent}
\bibinfo{author}{\bibfnamefont{J.-J.} \bibnamefont{Greffet}},
  \bibinfo{author}{\bibfnamefont{R.}~\bibnamefont{Carminati}},
  \bibinfo{author}{\bibfnamefont{K.}~\bibnamefont{Joulain}},
  \bibinfo{author}{\bibfnamefont{J.-P.} \bibnamefont{Mulet}},
  \bibinfo{author}{\bibfnamefont{S.}~\bibnamefont{Mainguy}}, \bibnamefont{and}
  \bibinfo{author}{\bibfnamefont{Y.}~\bibnamefont{Chen}},
  \bibinfo{journal}{Nature} \textbf{\bibinfo{volume}{416}}, \bibinfo{pages}{61}
  (\bibinfo{year}{2002}).

\bibitem[{\citenamefont{Marquier et~al.}(2004)\citenamefont{Marquier, Joulain,
  Mulet, Carminati, Greffet, and Chen}}]{marquier2004coherent}
\bibinfo{author}{\bibfnamefont{F.}~\bibnamefont{Marquier}},
  \bibinfo{author}{\bibfnamefont{K.}~\bibnamefont{Joulain}},
  \bibinfo{author}{\bibfnamefont{J.-P.} \bibnamefont{Mulet}},
  \bibinfo{author}{\bibfnamefont{R.}~\bibnamefont{Carminati}},
  \bibinfo{author}{\bibfnamefont{J.-J.} \bibnamefont{Greffet}},
  \bibnamefont{and} \bibinfo{author}{\bibfnamefont{Y.}~\bibnamefont{Chen}},
  \bibinfo{journal}{Physical Review B} \textbf{\bibinfo{volume}{69}},
  \bibinfo{pages}{155412} (\bibinfo{year}{2004}).

\bibitem[{\citenamefont{Joulain et~al.}(2005)\citenamefont{Joulain, Mulet,
  Marquier, Carminati, and Greffet}}]{Joulain05}
\bibinfo{author}{\bibfnamefont{K.}~\bibnamefont{Joulain}},
  \bibinfo{author}{\bibfnamefont{J.-P.} \bibnamefont{Mulet}},
  \bibinfo{author}{\bibfnamefont{F.}~\bibnamefont{Marquier}},
  \bibinfo{author}{\bibfnamefont{R.}~\bibnamefont{Carminati}},
  \bibnamefont{and} \bibinfo{author}{\bibfnamefont{J.-J.}
  \bibnamefont{Greffet}}, \bibinfo{journal}{Surf. Sci. Rep.}
  \textbf{\bibinfo{volume}{57}}, \bibinfo{pages}{59} (\bibinfo{year}{2005}).

\bibitem[{\citenamefont{Hesketh et~al.}(1988)\citenamefont{Hesketh, Zemel, and
  Gebhart}}]{hesketh1988polarized}
\bibinfo{author}{\bibfnamefont{P.~J.} \bibnamefont{Hesketh}},
  \bibinfo{author}{\bibfnamefont{J.~N.} \bibnamefont{Zemel}}, \bibnamefont{and}
  \bibinfo{author}{\bibfnamefont{B.}~\bibnamefont{Gebhart}},
  \bibinfo{journal}{Physical Review B} \textbf{\bibinfo{volume}{37}},
  \bibinfo{pages}{10803} (\bibinfo{year}{1988}).

\bibitem[{\citenamefont{Narayanaswamy and
  Chen}(2005)}]{narayanaswamy2005thermal}
\bibinfo{author}{\bibfnamefont{A.}~\bibnamefont{Narayanaswamy}}
  \bibnamefont{and} \bibinfo{author}{\bibfnamefont{G.}~\bibnamefont{Chen}},
  \bibinfo{journal}{Journal of Quantitative Spectroscopy and Radiative
  Transfer} \textbf{\bibinfo{volume}{93}}, \bibinfo{pages}{175}
  (\bibinfo{year}{2005}).

\bibitem[{\citenamefont{Marquier et~al.}(2015)\citenamefont{Marquier,
  Costantini, Lefebvre, Coutrot, Moldovan-Doyen, Hugonin, Boutami, Benisty, and
  Greffet}}]{marquier2015metallic}
\bibinfo{author}{\bibfnamefont{F.}~\bibnamefont{Marquier}},
  \bibinfo{author}{\bibfnamefont{D.}~\bibnamefont{Costantini}},
  \bibinfo{author}{\bibfnamefont{A.}~\bibnamefont{Lefebvre}},
  \bibinfo{author}{\bibfnamefont{A.-L.} \bibnamefont{Coutrot}},
  \bibinfo{author}{\bibfnamefont{I.}~\bibnamefont{Moldovan-Doyen}},
  \bibinfo{author}{\bibfnamefont{J.-P.} \bibnamefont{Hugonin}},
  \bibinfo{author}{\bibfnamefont{S.}~\bibnamefont{Boutami}},
  \bibinfo{author}{\bibfnamefont{H.}~\bibnamefont{Benisty}}, \bibnamefont{and}
  \bibinfo{author}{\bibfnamefont{J.-J.} \bibnamefont{Greffet}}, in
  \emph{\bibinfo{booktitle}{SPIE OPTO}} (\bibinfo{organization}{International
  Society for Optics and Photonics}, \bibinfo{year}{2015}), pp.
  \bibinfo{pages}{937004--937004}.

\bibitem[{\citenamefont{Kleiner et~al.}(2012)\citenamefont{Kleiner, Dahan,
  Frischwasser, and Hasman}}]{kleiner2012acrobatics}
\bibinfo{author}{\bibfnamefont{V.}~\bibnamefont{Kleiner}},
  \bibinfo{author}{\bibfnamefont{N.}~\bibnamefont{Dahan}},
  \bibinfo{author}{\bibfnamefont{K.}~\bibnamefont{Frischwasser}},
  \bibnamefont{and} \bibinfo{author}{\bibfnamefont{E.}~\bibnamefont{Hasman}},
  in \emph{\bibinfo{booktitle}{SPIE OPTO}}
  (\bibinfo{organization}{International Society for Optics and Photonics},
  \bibinfo{year}{2012}), pp. \bibinfo{pages}{82700R--82700R}.

\bibitem[{\citenamefont{Ribaudo et~al.}(2013)\citenamefont{Ribaudo, Peters,
  Ellis, Davids, and Shaner}}]{ribaudo2013highly}
\bibinfo{author}{\bibfnamefont{T.}~\bibnamefont{Ribaudo}},
  \bibinfo{author}{\bibfnamefont{D.~W.} \bibnamefont{Peters}},
  \bibinfo{author}{\bibfnamefont{A.~R.} \bibnamefont{Ellis}},
  \bibinfo{author}{\bibfnamefont{P.~S.} \bibnamefont{Davids}},
  \bibnamefont{and} \bibinfo{author}{\bibfnamefont{E.~A.}
  \bibnamefont{Shaner}}, \bibinfo{journal}{Optics express}
  \textbf{\bibinfo{volume}{21}}, \bibinfo{pages}{6837} (\bibinfo{year}{2013}).

\bibitem[{\citenamefont{Kollyukh et~al.}(2003)\citenamefont{Kollyukh, Liptuga,
  Morozhenko, and Pipa}}]{kollyukh2003thermal}
\bibinfo{author}{\bibfnamefont{O.}~\bibnamefont{Kollyukh}},
  \bibinfo{author}{\bibfnamefont{A.}~\bibnamefont{Liptuga}},
  \bibinfo{author}{\bibfnamefont{V.}~\bibnamefont{Morozhenko}},
  \bibnamefont{and} \bibinfo{author}{\bibfnamefont{V.}~\bibnamefont{Pipa}},
  \bibinfo{journal}{Optics Communications} \textbf{\bibinfo{volume}{225}},
  \bibinfo{pages}{349} (\bibinfo{year}{2003}).

\bibitem[{\citenamefont{Ben-Abdallah}(2004)}]{ben2004thermal}
\bibinfo{author}{\bibfnamefont{P.}~\bibnamefont{Ben-Abdallah}},
  \bibinfo{journal}{JOSA A} \textbf{\bibinfo{volume}{21}},
  \bibinfo{pages}{1368} (\bibinfo{year}{2004}).

\bibitem[{\citenamefont{Drevillon et~al.}(2011)\citenamefont{Drevillon,
  Joulain, Ben-Abdallah, and Nefzaoui}}]{drevillon2011far}
\bibinfo{author}{\bibfnamefont{J.}~\bibnamefont{Drevillon}},
  \bibinfo{author}{\bibfnamefont{K.}~\bibnamefont{Joulain}},
  \bibinfo{author}{\bibfnamefont{P.}~\bibnamefont{Ben-Abdallah}},
  \bibnamefont{and} \bibinfo{author}{\bibfnamefont{E.}~\bibnamefont{Nefzaoui}},
  \bibinfo{journal}{Journal of Applied Physics} \textbf{\bibinfo{volume}{109}},
  \bibinfo{pages}{034315} (\bibinfo{year}{2011}).

\bibitem[{\citenamefont{Wang et~al.}(2011)\citenamefont{Wang, Basu, and
  Zhang}}]{wang2011direct}
\bibinfo{author}{\bibfnamefont{L.}~\bibnamefont{Wang}},
  \bibinfo{author}{\bibfnamefont{S.}~\bibnamefont{Basu}}, \bibnamefont{and}
  \bibinfo{author}{\bibfnamefont{Z.}~\bibnamefont{Zhang}},
  \bibinfo{journal}{Journal of Heat Transfer} \textbf{\bibinfo{volume}{133}},
  \bibinfo{pages}{072701} (\bibinfo{year}{2011}).

\bibitem[{\citenamefont{Lee et~al.}(2008)\citenamefont{Lee, Wang, and
  Zhang}}]{lee2008coherent}
\bibinfo{author}{\bibfnamefont{B.}~\bibnamefont{Lee}},
  \bibinfo{author}{\bibfnamefont{L.}~\bibnamefont{Wang}}, \bibnamefont{and}
  \bibinfo{author}{\bibfnamefont{Z.}~\bibnamefont{Zhang}},
  \bibinfo{journal}{Optics Express} \textbf{\bibinfo{volume}{16}},
  \bibinfo{pages}{11328} (\bibinfo{year}{2008}).

\bibitem[{\citenamefont{Fu and Zhang}(2009)}]{fu2009thermal}
\bibinfo{author}{\bibfnamefont{C.}~\bibnamefont{Fu}} \bibnamefont{and}
  \bibinfo{author}{\bibfnamefont{Z.~M.} \bibnamefont{Zhang}},
  \bibinfo{journal}{Frontiers of Energy and Power Engineering in China}
  \textbf{\bibinfo{volume}{3}}, \bibinfo{pages}{11} (\bibinfo{year}{2009}).

\bibitem[{\citenamefont{Liu et~al.}(2011{\natexlab{a}})\citenamefont{Liu,
  Tyler, Starr, Starr, Jokerst, and Padilla}}]{Liu11}
\bibinfo{author}{\bibfnamefont{X.}~\bibnamefont{Liu}},
  \bibinfo{author}{\bibfnamefont{T.}~\bibnamefont{Tyler}},
  \bibinfo{author}{\bibfnamefont{T.}~\bibnamefont{Starr}},
  \bibinfo{author}{\bibfnamefont{A.~F.} \bibnamefont{Starr}},
  \bibinfo{author}{\bibfnamefont{N.~M.} \bibnamefont{Jokerst}},
  \bibnamefont{and} \bibinfo{author}{\bibfnamefont{W.~J.}
  \bibnamefont{Padilla}}, \bibinfo{journal}{Phys. Rev. Lett.}
  \textbf{\bibinfo{volume}{107}}, \bibinfo{pages}{045901}
  (\bibinfo{year}{2011}{\natexlab{a}}).

\bibitem[{\citenamefont{Liu et~al.}(2011{\natexlab{b}})\citenamefont{Liu,
  Tyler, Starr, Starr, Jokerst, and Padilla}}]{liu2011taming}
\bibinfo{author}{\bibfnamefont{X.}~\bibnamefont{Liu}},
  \bibinfo{author}{\bibfnamefont{T.}~\bibnamefont{Tyler}},
  \bibinfo{author}{\bibfnamefont{T.}~\bibnamefont{Starr}},
  \bibinfo{author}{\bibfnamefont{A.~F.} \bibnamefont{Starr}},
  \bibinfo{author}{\bibfnamefont{N.~M.} \bibnamefont{Jokerst}},
  \bibnamefont{and} \bibinfo{author}{\bibfnamefont{W.~J.}
  \bibnamefont{Padilla}}, \bibinfo{journal}{Physical review letters}
  \textbf{\bibinfo{volume}{107}}, \bibinfo{pages}{045901}
  (\bibinfo{year}{2011}{\natexlab{b}}).

\bibitem[{\citenamefont{Curto et~al.}(2010)\citenamefont{Curto, Volpe,
  Taminiau, Kreuzer, Quidant, and van Hulst}}]{curto2010unidirectional}
\bibinfo{author}{\bibfnamefont{A.~G.} \bibnamefont{Curto}},
  \bibinfo{author}{\bibfnamefont{G.}~\bibnamefont{Volpe}},
  \bibinfo{author}{\bibfnamefont{T.~H.} \bibnamefont{Taminiau}},
  \bibinfo{author}{\bibfnamefont{M.~P.} \bibnamefont{Kreuzer}},
  \bibinfo{author}{\bibfnamefont{R.}~\bibnamefont{Quidant}}, \bibnamefont{and}
  \bibinfo{author}{\bibfnamefont{N.~F.} \bibnamefont{van Hulst}},
  \bibinfo{journal}{Science} \textbf{\bibinfo{volume}{329}},
  \bibinfo{pages}{930} (\bibinfo{year}{2010}).

\bibitem[{\citenamefont{Kosako et~al.}(2010)\citenamefont{Kosako, Kadoya, and
  Hofmann}}]{kosako2010directional}
\bibinfo{author}{\bibfnamefont{T.}~\bibnamefont{Kosako}},
  \bibinfo{author}{\bibfnamefont{Y.}~\bibnamefont{Kadoya}}, \bibnamefont{and}
  \bibinfo{author}{\bibfnamefont{H.~F.} \bibnamefont{Hofmann}},
  \bibinfo{journal}{Nature Photonics} \textbf{\bibinfo{volume}{4}},
  \bibinfo{pages}{312} (\bibinfo{year}{2010}).

\bibitem[{\citenamefont{Teperik and Degiron}(2011)}]{teperik2011numerical}
\bibinfo{author}{\bibfnamefont{T.}~\bibnamefont{Teperik}} \bibnamefont{and}
  \bibinfo{author}{\bibfnamefont{A.}~\bibnamefont{Degiron}},
  \bibinfo{journal}{Physical Review B} \textbf{\bibinfo{volume}{83}},
  \bibinfo{pages}{245408} (\bibinfo{year}{2011}).

\bibitem[{\citenamefont{Thomas et~al.}(2004)\citenamefont{Thomas, Greffet,
  Carminati, and Arias-Gonzalez}}]{thomas2004single}
\bibinfo{author}{\bibfnamefont{M.}~\bibnamefont{Thomas}},
  \bibinfo{author}{\bibfnamefont{J.-J.} \bibnamefont{Greffet}},
  \bibinfo{author}{\bibfnamefont{R.}~\bibnamefont{Carminati}},
  \bibnamefont{and}
  \bibinfo{author}{\bibfnamefont{J.}~\bibnamefont{Arias-Gonzalez}},
  \bibinfo{journal}{Applied physics letters} \textbf{\bibinfo{volume}{85}},
  \bibinfo{pages}{3863} (\bibinfo{year}{2004}).

\bibitem[{\citenamefont{Li et~al.}(2007)\citenamefont{Li, Kattawar, You, Zhai,
  and Yang}}]{li2007fdtd}
\bibinfo{author}{\bibfnamefont{C.}~\bibnamefont{Li}},
  \bibinfo{author}{\bibfnamefont{G.~W.} \bibnamefont{Kattawar}},
  \bibinfo{author}{\bibfnamefont{Y.}~\bibnamefont{You}},
  \bibinfo{author}{\bibfnamefont{P.}~\bibnamefont{Zhai}}, \bibnamefont{and}
  \bibinfo{author}{\bibfnamefont{P.}~\bibnamefont{Yang}},
  \bibinfo{journal}{Journal of Quantitative Spectroscopy and Radiative
  Transfer} \textbf{\bibinfo{volume}{106}}, \bibinfo{pages}{257}
  (\bibinfo{year}{2007}).

\bibitem[{\citenamefont{Vandenbem et~al.}(2009)\citenamefont{Vandenbem,
  Froufe-P{\'e}rez, and Carminati}}]{vandenbem2009fluorescence}
\bibinfo{author}{\bibfnamefont{C.}~\bibnamefont{Vandenbem}},
  \bibinfo{author}{\bibfnamefont{L.}~\bibnamefont{Froufe-P{\'e}rez}},
  \bibnamefont{and}
  \bibinfo{author}{\bibfnamefont{R.}~\bibnamefont{Carminati}},
  \bibinfo{journal}{Journal of Optics A: Pure and Applied Optics}
  \textbf{\bibinfo{volume}{11}}, \bibinfo{pages}{114007}
  (\bibinfo{year}{2009}).

\bibitem[{\citenamefont{Vandenbem et~al.}(2010)\citenamefont{Vandenbem, Brayer,
  Froufe-P{\'e}rez, and Carminati}}]{vandenbem2010controlling}
\bibinfo{author}{\bibfnamefont{C.}~\bibnamefont{Vandenbem}},
  \bibinfo{author}{\bibfnamefont{D.}~\bibnamefont{Brayer}},
  \bibinfo{author}{\bibfnamefont{L.}~\bibnamefont{Froufe-P{\'e}rez}},
  \bibnamefont{and}
  \bibinfo{author}{\bibfnamefont{R.}~\bibnamefont{Carminati}},
  \bibinfo{journal}{Physical Review B} \textbf{\bibinfo{volume}{81}},
  \bibinfo{pages}{085444} (\bibinfo{year}{2010}).

\bibitem[{\citenamefont{Mohammadi et~al.}(2008)\citenamefont{Mohammadi,
  Sandoghdar, and Agio}}]{mohammadi2008gold}
\bibinfo{author}{\bibfnamefont{A.}~\bibnamefont{Mohammadi}},
  \bibinfo{author}{\bibfnamefont{V.}~\bibnamefont{Sandoghdar}},
  \bibnamefont{and} \bibinfo{author}{\bibfnamefont{M.}~\bibnamefont{Agio}},
  \bibinfo{journal}{New Journal of Physics} \textbf{\bibinfo{volume}{10}},
  \bibinfo{pages}{105015} (\bibinfo{year}{2008}).

\bibitem[{\citenamefont{Schuller et~al.}(2009)\citenamefont{Schuller, Taubner,
  and Brongersma}}]{schuller2009optical}
\bibinfo{author}{\bibfnamefont{J.~A.} \bibnamefont{Schuller}},
  \bibinfo{author}{\bibfnamefont{T.}~\bibnamefont{Taubner}}, \bibnamefont{and}
  \bibinfo{author}{\bibfnamefont{M.~L.} \bibnamefont{Brongersma}},
  \bibinfo{journal}{Nature Photonics} \textbf{\bibinfo{volume}{3}},
  \bibinfo{pages}{658} (\bibinfo{year}{2009}).

\bibitem[{\citenamefont{Novotny and Van~Hulst}(2011)}]{novotny2011antennas}
\bibinfo{author}{\bibfnamefont{L.}~\bibnamefont{Novotny}} \bibnamefont{and}
  \bibinfo{author}{\bibfnamefont{N.}~\bibnamefont{Van~Hulst}},
  \bibinfo{journal}{Nature Photonics} \textbf{\bibinfo{volume}{5}},
  \bibinfo{pages}{83} (\bibinfo{year}{2011}).

\bibitem[{\citenamefont{Balanis}(2005)}]{balanis2005antenna}
\bibinfo{author}{\bibfnamefont{C.~A.} \bibnamefont{Balanis}},
  \emph{\bibinfo{title}{Antenna theory: analysis and design}},
  vol.~\bibinfo{volume}{1} (\bibinfo{publisher}{John Wiley \& Sons},
  \bibinfo{year}{2005}).

\bibitem[{\citenamefont{Taminiau et~al.}(2007)\citenamefont{Taminiau, Moerland,
  Segerink, Kuipers, and van Hulst}}]{taminiau2007lambda}
\bibinfo{author}{\bibfnamefont{T.~H.} \bibnamefont{Taminiau}},
  \bibinfo{author}{\bibfnamefont{R.~J.} \bibnamefont{Moerland}},
  \bibinfo{author}{\bibfnamefont{F.~B.} \bibnamefont{Segerink}},
  \bibinfo{author}{\bibfnamefont{L.}~\bibnamefont{Kuipers}}, \bibnamefont{and}
  \bibinfo{author}{\bibfnamefont{N.~F.} \bibnamefont{van Hulst}},
  \bibinfo{journal}{Nano letters} \textbf{\bibinfo{volume}{7}},
  \bibinfo{pages}{28} (\bibinfo{year}{2007}).

\bibitem[{\citenamefont{Alavi~Lavasani and Pakizeh}(2012)}]{alavi2012color}
\bibinfo{author}{\bibfnamefont{S.}~\bibnamefont{Alavi~Lavasani}}
  \bibnamefont{and} \bibinfo{author}{\bibfnamefont{T.}~\bibnamefont{Pakizeh}},
  \bibinfo{journal}{JOSA B} \textbf{\bibinfo{volume}{29}},
  \bibinfo{pages}{1361} (\bibinfo{year}{2012}).

\bibitem[{\citenamefont{Shegai et~al.}(2011)\citenamefont{Shegai, Chen,
  Miljkovi{\'c}, Zengin, Johansson, and K{\"a}ll}}]{shegai2011bimetallic}
\bibinfo{author}{\bibfnamefont{T.}~\bibnamefont{Shegai}},
  \bibinfo{author}{\bibfnamefont{S.}~\bibnamefont{Chen}},
  \bibinfo{author}{\bibfnamefont{V.~D.} \bibnamefont{Miljkovi{\'c}}},
  \bibinfo{author}{\bibfnamefont{G.}~\bibnamefont{Zengin}},
  \bibinfo{author}{\bibfnamefont{P.}~\bibnamefont{Johansson}},
  \bibnamefont{and} \bibinfo{author}{\bibfnamefont{M.}~\bibnamefont{K{\"a}ll}},
  \bibinfo{journal}{Nature communications} \textbf{\bibinfo{volume}{2}},
  \bibinfo{pages}{481} (\bibinfo{year}{2011}).

\bibitem[{\citenamefont{Balandin}(2011)}]{balandin2011thermal}
\bibinfo{author}{\bibfnamefont{A.~A.} \bibnamefont{Balandin}},
  \bibinfo{journal}{Nature materials} \textbf{\bibinfo{volume}{10}},
  \bibinfo{pages}{569} (\bibinfo{year}{2011}).

\bibitem[{\citenamefont{Reifenberg et~al.}(2008)\citenamefont{Reifenberg,
  Kencke, and Goodson}}]{reifenberg2008impact}
\bibinfo{author}{\bibfnamefont{J.~P.} \bibnamefont{Reifenberg}},
  \bibinfo{author}{\bibfnamefont{D.~L.} \bibnamefont{Kencke}},
  \bibnamefont{and} \bibinfo{author}{\bibfnamefont{K.~E.}
  \bibnamefont{Goodson}}, \bibinfo{journal}{Electron Device Letters, IEEE}
  \textbf{\bibinfo{volume}{29}}, \bibinfo{pages}{1112} (\bibinfo{year}{2008}).

\bibitem[{\citenamefont{Marconnet et~al.}(2013)\citenamefont{Marconnet, Panzer,
  and Goodson}}]{marconnet2013thermal}
\bibinfo{author}{\bibfnamefont{A.~M.} \bibnamefont{Marconnet}},
  \bibinfo{author}{\bibfnamefont{M.~A.} \bibnamefont{Panzer}},
  \bibnamefont{and} \bibinfo{author}{\bibfnamefont{K.~E.}
  \bibnamefont{Goodson}}, \bibinfo{journal}{Reviews of Modern Physics}
  \textbf{\bibinfo{volume}{85}}, \bibinfo{pages}{1295} (\bibinfo{year}{2013}).

\bibitem[{\citenamefont{Stevens et~al.}(2007)\citenamefont{Stevens, Zhigilei,
  and Norris}}]{stevens2007effects}
\bibinfo{author}{\bibfnamefont{R.~J.} \bibnamefont{Stevens}},
  \bibinfo{author}{\bibfnamefont{L.~V.} \bibnamefont{Zhigilei}},
  \bibnamefont{and} \bibinfo{author}{\bibfnamefont{P.~M.}
  \bibnamefont{Norris}}, \bibinfo{journal}{International Journal of Heat and
  Mass Transfer} \textbf{\bibinfo{volume}{50}}, \bibinfo{pages}{3977}
  (\bibinfo{year}{2007}).

\bibitem[{\citenamefont{Merabia et~al.}(2009)\citenamefont{Merabia, Keblinski,
  Joly, Lewis, and Barrat}}]{merabia2009critical}
\bibinfo{author}{\bibfnamefont{S.}~\bibnamefont{Merabia}},
  \bibinfo{author}{\bibfnamefont{P.}~\bibnamefont{Keblinski}},
  \bibinfo{author}{\bibfnamefont{L.}~\bibnamefont{Joly}},
  \bibinfo{author}{\bibfnamefont{L.~J.} \bibnamefont{Lewis}}, \bibnamefont{and}
  \bibinfo{author}{\bibfnamefont{J.-L.} \bibnamefont{Barrat}},
  \bibinfo{journal}{PRE} \textbf{\bibinfo{volume}{79}}, \bibinfo{pages}{021404}
  (\bibinfo{year}{2009}).

\bibitem[{\citenamefont{Islam et~al.}(2013)\citenamefont{Islam, Li, Dorgan,
  Bae, and Pop}}]{islam2013role}
\bibinfo{author}{\bibfnamefont{S.}~\bibnamefont{Islam}},
  \bibinfo{author}{\bibfnamefont{Z.}~\bibnamefont{Li}},
  \bibinfo{author}{\bibfnamefont{V.~E.} \bibnamefont{Dorgan}},
  \bibinfo{author}{\bibfnamefont{M.-H.} \bibnamefont{Bae}}, \bibnamefont{and}
  \bibinfo{author}{\bibfnamefont{E.}~\bibnamefont{Pop}},
  \bibinfo{journal}{Electron Device Letters, IEEE}
  \textbf{\bibinfo{volume}{34}}, \bibinfo{pages}{166} (\bibinfo{year}{2013}).

\bibitem[{\citenamefont{Yeo et~al.}(2014)\citenamefont{Yeo, Kim, Hong, Lee,
  Kwon, Lee, Park, Manoroktul, Lee, Lee et~al.}}]{yeo2014single}
\bibinfo{author}{\bibfnamefont{J.}~\bibnamefont{Yeo}},
  \bibinfo{author}{\bibfnamefont{G.}~\bibnamefont{Kim}},
  \bibinfo{author}{\bibfnamefont{S.}~\bibnamefont{Hong}},
  \bibinfo{author}{\bibfnamefont{J.}~\bibnamefont{Lee}},
  \bibinfo{author}{\bibfnamefont{J.}~\bibnamefont{Kwon}},
  \bibinfo{author}{\bibfnamefont{H.}~\bibnamefont{Lee}},
  \bibinfo{author}{\bibfnamefont{H.}~\bibnamefont{Park}},
  \bibinfo{author}{\bibfnamefont{W.}~\bibnamefont{Manoroktul}},
  \bibinfo{author}{\bibfnamefont{M.-T.} \bibnamefont{Lee}},
  \bibinfo{author}{\bibfnamefont{B.~J.} \bibnamefont{Lee}},
  \bibnamefont{et~al.}, \bibinfo{journal}{Small} \textbf{\bibinfo{volume}{10}},
  \bibinfo{pages}{5015} (\bibinfo{year}{2014}).

\bibitem[{\citenamefont{King et~al.}(2013)\citenamefont{King, Bhatia, Felts,
  Kim, Kwon, Lee, Somnath, and Rosenberger}}]{king2013heated}
\bibinfo{author}{\bibfnamefont{W.~P.} \bibnamefont{King}},
  \bibinfo{author}{\bibfnamefont{B.}~\bibnamefont{Bhatia}},
  \bibinfo{author}{\bibfnamefont{J.~R.} \bibnamefont{Felts}},
  \bibinfo{author}{\bibfnamefont{H.~J.} \bibnamefont{Kim}},
  \bibinfo{author}{\bibfnamefont{B.}~\bibnamefont{Kwon}},
  \bibinfo{author}{\bibfnamefont{B.}~\bibnamefont{Lee}},
  \bibinfo{author}{\bibfnamefont{S.}~\bibnamefont{Somnath}}, \bibnamefont{and}
  \bibinfo{author}{\bibfnamefont{M.}~\bibnamefont{Rosenberger}},
  \bibinfo{journal}{Annual Review of Heat Transfer}
  \textbf{\bibinfo{volume}{16}} (\bibinfo{year}{2013}).

\bibitem[{\citenamefont{Petit-Watelot et~al.}(2012)\citenamefont{Petit-Watelot,
  Otxoa, Manfrini, Van~Roy, Lagae, Kim, and Devolder}}]{petit2012understanding}
\bibinfo{author}{\bibfnamefont{S.}~\bibnamefont{Petit-Watelot}},
  \bibinfo{author}{\bibfnamefont{R.~M.} \bibnamefont{Otxoa}},
  \bibinfo{author}{\bibfnamefont{M.}~\bibnamefont{Manfrini}},
  \bibinfo{author}{\bibfnamefont{W.}~\bibnamefont{Van~Roy}},
  \bibinfo{author}{\bibfnamefont{L.}~\bibnamefont{Lagae}},
  \bibinfo{author}{\bibfnamefont{J.-V.} \bibnamefont{Kim}}, \bibnamefont{and}
  \bibinfo{author}{\bibfnamefont{T.}~\bibnamefont{Devolder}},
  \bibinfo{journal}{Physical review letters} \textbf{\bibinfo{volume}{109}},
  \bibinfo{pages}{267205} (\bibinfo{year}{2012}).

\bibitem[{\citenamefont{Dombrovsky}(2000)}]{dombrovsky2000thermal}
\bibinfo{author}{\bibfnamefont{L.~A.} \bibnamefont{Dombrovsky}},
  \bibinfo{journal}{International Journal of Heat and Mass Transfer}
  \textbf{\bibinfo{volume}{43}}, \bibinfo{pages}{1661} (\bibinfo{year}{2000}).

\bibitem[{\citenamefont{Luo et~al.}(2004)\citenamefont{Luo, Narayanaswamy,
  Ghen, and Joannopoulos}}]{Luo04:thermal}
\bibinfo{author}{\bibfnamefont{C.}~\bibnamefont{Luo}},
  \bibinfo{author}{\bibfnamefont{A.}~\bibnamefont{Narayanaswamy}},
  \bibinfo{author}{\bibfnamefont{G.}~\bibnamefont{Ghen}}, \bibnamefont{and}
  \bibinfo{author}{\bibfnamefont{J.~D.} \bibnamefont{Joannopoulos}},
  \bibinfo{journal}{Phys. Rev. Lett.} \textbf{\bibinfo{volume}{93}},
  \bibinfo{pages}{213905} (\bibinfo{year}{2004}).

\bibitem[{\citenamefont{Polimeridis et~al.}(2015)\citenamefont{Polimeridis,
  Reid, Jin, Johnson, White, and Rodriguezz}}]{polimeridis2015fluctuating}
\bibinfo{author}{\bibfnamefont{A.~G.} \bibnamefont{Polimeridis}},
  \bibinfo{author}{\bibfnamefont{M.}~\bibnamefont{Reid}},
  \bibinfo{author}{\bibfnamefont{W.}~\bibnamefont{Jin}},
  \bibinfo{author}{\bibfnamefont{S.~G.} \bibnamefont{Johnson}},
  \bibinfo{author}{\bibfnamefont{J.~K.} \bibnamefont{White}}, \bibnamefont{and}
  \bibinfo{author}{\bibfnamefont{A.~W.} \bibnamefont{Rodriguezz}},
  \bibinfo{journal}{arXiv preprint arXiv:1505.05026}  (\bibinfo{year}{2015}).

\bibitem[{\citenamefont{Polimeridis et~al.}(2014)\citenamefont{Polimeridis,
  Villena, Daniel, and White}}]{polimeridis2014stable}
\bibinfo{author}{\bibfnamefont{A.}~\bibnamefont{Polimeridis}},
  \bibinfo{author}{\bibfnamefont{J.}~\bibnamefont{Villena}},
  \bibinfo{author}{\bibfnamefont{L.}~\bibnamefont{Daniel}}, \bibnamefont{and}
  \bibinfo{author}{\bibfnamefont{J.}~\bibnamefont{White}},
  \bibinfo{journal}{Journal of Computational Physics}
  \textbf{\bibinfo{volume}{269}}, \bibinfo{pages}{280} (\bibinfo{year}{2014}).

\bibitem[{\citenamefont{Landau et~al.}(1960)\citenamefont{Landau, Lifshitz, and
  Pitaevski{\u{\i}}}}]{landau:stat2}
\bibinfo{author}{\bibfnamefont{L.~D.} \bibnamefont{Landau}},
  \bibinfo{author}{\bibfnamefont{E.~M.} \bibnamefont{Lifshitz}},
  \bibnamefont{and} \bibinfo{author}{\bibfnamefont{L.~P.}
  \bibnamefont{Pitaevski{\u{\i}}}}, \emph{\bibinfo{title}{Statistical Physics
  Part 2}}, vol.~\bibinfo{volume}{9} (\bibinfo{publisher}{Pergamon},
  \bibinfo{address}{Oxford}, \bibinfo{year}{1960}).

\bibitem[{\citenamefont{Rodriguez et~al.}(2013)\citenamefont{Rodriguez, Reid,
  and Johnson}}]{rodriguez2013fluctuating}
\bibinfo{author}{\bibfnamefont{A.~W.} \bibnamefont{Rodriguez}},
  \bibinfo{author}{\bibfnamefont{M.~H.} \bibnamefont{Reid}}, \bibnamefont{and}
  \bibinfo{author}{\bibfnamefont{S.~G.} \bibnamefont{Johnson}},
  \bibinfo{journal}{Physical Review B} \textbf{\bibinfo{volume}{88}},
  \bibinfo{pages}{054305} (\bibinfo{year}{2013}).

\bibitem[{\citenamefont{Johnson}(2011)}]{Johnson11:review}
\bibinfo{author}{\bibfnamefont{S.~G.} \bibnamefont{Johnson}}, in
  \emph{\bibinfo{booktitle}{Casimir Physics}}, edited by
  \bibinfo{editor}{\bibfnamefont{D.~A.~R.} \bibnamefont{Dalvit}},
  \bibinfo{editor}{\bibfnamefont{P.}~\bibnamefont{Milonni}},
  \bibinfo{editor}{\bibfnamefont{D.}~\bibnamefont{Roberts}}, \bibnamefont{and}
  \bibinfo{editor}{\bibfnamefont{F.~d.} \bibnamefont{Rosa}}
  (\bibinfo{publisher}{Springer--Verlag}, \bibinfo{year}{2011}), vol.
  \bibinfo{volume}{836} of \emph{\bibinfo{series}{Lecture Notes in Physics}},
  chap.~\bibinfo{chapter}{6}, pp. \bibinfo{pages}{175--218}.

\bibitem[{\citenamefont{Xiong et~al.}(2009)\citenamefont{Xiong, Liao, and
  Pop}}]{xiong2009inducing}
\bibinfo{author}{\bibfnamefont{F.}~\bibnamefont{Xiong}},
  \bibinfo{author}{\bibfnamefont{A.}~\bibnamefont{Liao}}, \bibnamefont{and}
  \bibinfo{author}{\bibfnamefont{E.}~\bibnamefont{Pop}},
  \bibinfo{journal}{Applied Physics Letters} \textbf{\bibinfo{volume}{95}},
  \bibinfo{pages}{243103} (\bibinfo{year}{2009}).

\bibitem[{\citenamefont{Liang et~al.}(2012)\citenamefont{Liang, Jeyasingh,
  Chen, and Wong}}]{liang2012ultra}
\bibinfo{author}{\bibfnamefont{J.}~\bibnamefont{Liang}},
  \bibinfo{author}{\bibfnamefont{R.~G.~D.} \bibnamefont{Jeyasingh}},
  \bibinfo{author}{\bibfnamefont{H.-Y.} \bibnamefont{Chen}}, \bibnamefont{and}
  \bibinfo{author}{\bibfnamefont{H.}~\bibnamefont{Wong}},
  \bibinfo{journal}{Electron Devices, IEEE Transactions on}
  \textbf{\bibinfo{volume}{59}}, \bibinfo{pages}{1155} (\bibinfo{year}{2012}).

\bibitem[{Note1()}]{Note1}
Note1, \bibinfo{note}{note that at these temperatures convective and radiative
  effects are negligible compared to conductive transfer, allowing us to
  consider the radiation and conduction problems separately.}

\bibitem[{\citenamefont{Lyeo et~al.}(2006)\citenamefont{Lyeo, Cahill, Lee,
  Abelson, Kwon, Kim, Bishop, and Cheong}}]{lyeo2006thermal}
\bibinfo{author}{\bibfnamefont{H.-K.} \bibnamefont{Lyeo}},
  \bibinfo{author}{\bibfnamefont{D.~G.} \bibnamefont{Cahill}},
  \bibinfo{author}{\bibfnamefont{B.-S.} \bibnamefont{Lee}},
  \bibinfo{author}{\bibfnamefont{J.~R.} \bibnamefont{Abelson}},
  \bibinfo{author}{\bibfnamefont{M.-H.} \bibnamefont{Kwon}},
  \bibinfo{author}{\bibfnamefont{K.-B.} \bibnamefont{Kim}},
  \bibinfo{author}{\bibfnamefont{S.~G.} \bibnamefont{Bishop}},
  \bibnamefont{and} \bibinfo{author}{\bibfnamefont{B.-k.}
  \bibnamefont{Cheong}}, \bibinfo{journal}{Applied Physics Letters}
  \textbf{\bibinfo{volume}{89}}, \bibinfo{pages}{151904}
  (\bibinfo{year}{2006}).

\bibitem[{\citenamefont{Tsafack et~al.}(2011)\citenamefont{Tsafack, Piccinini,
  Lee, Pop, and Rudan}}]{tsafack2011electronic}
\bibinfo{author}{\bibfnamefont{T.}~\bibnamefont{Tsafack}},
  \bibinfo{author}{\bibfnamefont{E.}~\bibnamefont{Piccinini}},
  \bibinfo{author}{\bibfnamefont{B.-S.} \bibnamefont{Lee}},
  \bibinfo{author}{\bibfnamefont{E.}~\bibnamefont{Pop}}, \bibnamefont{and}
  \bibinfo{author}{\bibfnamefont{M.}~\bibnamefont{Rudan}},
  \bibinfo{journal}{Journal of Applied Physics} \textbf{\bibinfo{volume}{110}},
  \bibinfo{pages}{063716} (\bibinfo{year}{2011}).

\bibitem[{\citenamefont{Shportko et~al.}(2008)\citenamefont{Shportko, Kremers,
  Woda, Lencer, Robertson, and Wuttig}}]{shportko2008resonant}
\bibinfo{author}{\bibfnamefont{K.}~\bibnamefont{Shportko}},
  \bibinfo{author}{\bibfnamefont{S.}~\bibnamefont{Kremers}},
  \bibinfo{author}{\bibfnamefont{M.}~\bibnamefont{Woda}},
  \bibinfo{author}{\bibfnamefont{D.}~\bibnamefont{Lencer}},
  \bibinfo{author}{\bibfnamefont{J.}~\bibnamefont{Robertson}},
  \bibnamefont{and} \bibinfo{author}{\bibfnamefont{M.}~\bibnamefont{Wuttig}},
  \bibinfo{journal}{Nature materials} \textbf{\bibinfo{volume}{7}},
  \bibinfo{pages}{653} (\bibinfo{year}{2008}).

\bibitem[{\citenamefont{Li et~al.}(2008)\citenamefont{Li, Choi, Byun, Kim, Sim,
  and Kim}}]{li2008high}
\bibinfo{author}{\bibfnamefont{X.~Z.} \bibnamefont{Li}},
  \bibinfo{author}{\bibfnamefont{J.~K.} \bibnamefont{Choi}},
  \bibinfo{author}{\bibfnamefont{Y.~S.} \bibnamefont{Byun}},
  \bibinfo{author}{\bibfnamefont{S.~Y.} \bibnamefont{Kim}},
  \bibinfo{author}{\bibfnamefont{K.~S.} \bibnamefont{Sim}}, \bibnamefont{and}
  \bibinfo{author}{\bibfnamefont{S.~K.} \bibnamefont{Kim}},
  \bibinfo{journal}{Japanese Journal of Applied Physics}
  \textbf{\bibinfo{volume}{47}}, \bibinfo{pages}{5477} (\bibinfo{year}{2008}).

\bibitem[{\citenamefont{Mash and Motulevich}(1973)}]{mash1973optical}
\bibinfo{author}{\bibfnamefont{I.}~\bibnamefont{Mash}} \bibnamefont{and}
  \bibinfo{author}{\bibfnamefont{G.}~\bibnamefont{Motulevich}},
  \bibinfo{journal}{SOVIET PHYSICS JETP} \textbf{\bibinfo{volume}{36}}
  (\bibinfo{year}{1973}).

\bibitem[{\citenamefont{Kischkat et~al.}(2012)\citenamefont{Kischkat, Peters,
  Gruska, Semtsiv, Chashnikova, Klinkm{\"u}ller, Fedosenko, Machulik,
  Aleksandrova, Monastyrskyi et~al.}}]{kischkat2012mid}
\bibinfo{author}{\bibfnamefont{J.}~\bibnamefont{Kischkat}},
  \bibinfo{author}{\bibfnamefont{S.}~\bibnamefont{Peters}},
  \bibinfo{author}{\bibfnamefont{B.}~\bibnamefont{Gruska}},
  \bibinfo{author}{\bibfnamefont{M.}~\bibnamefont{Semtsiv}},
  \bibinfo{author}{\bibfnamefont{M.}~\bibnamefont{Chashnikova}},
  \bibinfo{author}{\bibfnamefont{M.}~\bibnamefont{Klinkm{\"u}ller}},
  \bibinfo{author}{\bibfnamefont{O.}~\bibnamefont{Fedosenko}},
  \bibinfo{author}{\bibfnamefont{S.}~\bibnamefont{Machulik}},
  \bibinfo{author}{\bibfnamefont{A.}~\bibnamefont{Aleksandrova}},
  \bibinfo{author}{\bibfnamefont{G.}~\bibnamefont{Monastyrskyi}},
  \bibnamefont{et~al.}, \bibinfo{journal}{Applied optics}
  \textbf{\bibinfo{volume}{51}}, \bibinfo{pages}{6789} (\bibinfo{year}{2012}).

\bibitem[{\citenamefont{Bohren and Huffman}(2008)}]{bohren2008absorption}
\bibinfo{author}{\bibfnamefont{C.~F.} \bibnamefont{Bohren}} \bibnamefont{and}
  \bibinfo{author}{\bibfnamefont{D.~R.} \bibnamefont{Huffman}},
  \emph{\bibinfo{title}{Absorption and scattering of light by small particles}}
  (\bibinfo{publisher}{John Wiley \& Sons}, \bibinfo{year}{2008}).

\bibitem[{\citenamefont{Yu et~al.}(2013)\citenamefont{Yu, Sergeant, Skauli,
  Zhang, Wang, and Fan}}]{yu2013enhancing}
\bibinfo{author}{\bibfnamefont{Z.}~\bibnamefont{Yu}},
  \bibinfo{author}{\bibfnamefont{N.~P.} \bibnamefont{Sergeant}},
  \bibinfo{author}{\bibfnamefont{T.}~\bibnamefont{Skauli}},
  \bibinfo{author}{\bibfnamefont{G.}~\bibnamefont{Zhang}},
  \bibinfo{author}{\bibfnamefont{H.}~\bibnamefont{Wang}}, \bibnamefont{and}
  \bibinfo{author}{\bibfnamefont{S.}~\bibnamefont{Fan}},
  \bibinfo{journal}{Nature communications} \textbf{\bibinfo{volume}{4}},
  \bibinfo{pages}{1730} (\bibinfo{year}{2013}).

\bibitem[{\citenamefont{Weinstein}(1960)}]{weinstein1960validity}
\bibinfo{author}{\bibfnamefont{M.}~\bibnamefont{Weinstein}},
  \bibinfo{journal}{American Journal of Physics} \textbf{\bibinfo{volume}{28}},
  \bibinfo{pages}{123} (\bibinfo{year}{1960}).

\bibitem[{\citenamefont{Fischer and Fellmuth}(2005)}]{fischer2005temperature}
\bibinfo{author}{\bibfnamefont{J.}~\bibnamefont{Fischer}} \bibnamefont{and}
  \bibinfo{author}{\bibfnamefont{B.}~\bibnamefont{Fellmuth}},
  \bibinfo{journal}{Reports on progress in physics}
  \textbf{\bibinfo{volume}{68}}, \bibinfo{pages}{1043} (\bibinfo{year}{2005}).

\bibitem[{\citenamefont{Banaei and Abouraddy}(2013)}]{banaei2013design}
\bibinfo{author}{\bibfnamefont{E.-H.} \bibnamefont{Banaei}} \bibnamefont{and}
  \bibinfo{author}{\bibfnamefont{A.~F.} \bibnamefont{Abouraddy}},
  \bibinfo{journal}{Progress in Photovoltaics: Research and Applications}
  (\bibinfo{year}{2013}).

\bibitem[{\citenamefont{Le~Ru and Etchegoin}(2008)}]{LeRu08}
\bibinfo{author}{\bibfnamefont{E.}~\bibnamefont{Le~Ru}} \bibnamefont{and}
  \bibinfo{author}{\bibfnamefont{P.}~\bibnamefont{Etchegoin}},
  \emph{\bibinfo{title}{Principles of Surface-Enhanced Raman Spectroscopy and
  related plasmonic effects}} (\bibinfo{publisher}{Elsevier Science},
  \bibinfo{year}{2008}).

\bibitem[{\citenamefont{Rud{\'e} et~al.}(2013)\citenamefont{Rud{\'e}, Pello,
  Simpson, Osmond, Roelkens, van~der Tol, and Pruneri}}]{rude2013optical}
\bibinfo{author}{\bibfnamefont{M.}~\bibnamefont{Rud{\'e}}},
  \bibinfo{author}{\bibfnamefont{J.}~\bibnamefont{Pello}},
  \bibinfo{author}{\bibfnamefont{R.~E.} \bibnamefont{Simpson}},
  \bibinfo{author}{\bibfnamefont{J.}~\bibnamefont{Osmond}},
  \bibinfo{author}{\bibfnamefont{G.}~\bibnamefont{Roelkens}},
  \bibinfo{author}{\bibfnamefont{J.~J.} \bibnamefont{van~der Tol}},
  \bibnamefont{and} \bibinfo{author}{\bibfnamefont{V.}~\bibnamefont{Pruneri}},
  \bibinfo{journal}{Applied Physics Letters} \textbf{\bibinfo{volume}{103}},
  \bibinfo{pages}{141119} (\bibinfo{year}{2013}).

\bibitem[{\citenamefont{Kim et~al.}(2012)\citenamefont{Kim, Naik, Emani, and
  Boltasseva}}]{kim2012plasmonic}
\bibinfo{author}{\bibfnamefont{J.}~\bibnamefont{Kim}},
  \bibinfo{author}{\bibfnamefont{G.~V.} \bibnamefont{Naik}},
  \bibinfo{author}{\bibfnamefont{N.~K.} \bibnamefont{Emani}}, \bibnamefont{and}
  \bibinfo{author}{\bibfnamefont{A.}~\bibnamefont{Boltasseva}},
  \bibinfo{journal}{arXiv preprint arXiv:1211.5988}  (\bibinfo{year}{2012}).

\bibitem[{\citenamefont{Shahil and Balandin}(2012)}]{shahil2012thermal}
\bibinfo{author}{\bibfnamefont{K.~M.} \bibnamefont{Shahil}} \bibnamefont{and}
  \bibinfo{author}{\bibfnamefont{A.~A.} \bibnamefont{Balandin}},
  \bibinfo{journal}{Solid State Communications} \textbf{\bibinfo{volume}{152}},
  \bibinfo{pages}{1331} (\bibinfo{year}{2012}).

\end{thebibliography}
\end{document}